\documentclass{aastex63}

\newcommand{\kms}{km s$^{-1}$}
\newcommand{\msun}{$M_{\sun}$}
\newcommand{\mJybeam}{mJy\thinspace beam$^{-1}$}
\newcommand{\Jybeam}{Jy\thinspace beam$^{-1}$}

\newcommand{\sixcm}{$\lambda 6$ cm}
\newcommand{\HI}{\ion{H}{1}}
\newcommand{\Halpha}{H$\alpha$}

\newcommand{\HII}{\ion{H}{2}}
\newcommand{\Hmolec}{H$_2 \thinspace 1\rightarrow 0$ S(1)}
\newcommand{\CO}{$^{12}$CO $J=1\rightarrow 0$}

\newcommand{\Jybeamkms}{Jy\thinspace beam$^{-1}$ km\thinspace s$^{-1}$}
\newcommand{\msunpc}{$M_{\sun}$ pc$^{-2}$}

\received{January 28, 2019}
\revised{February 25 2020}
\accepted{February 29, 2020}
\submitjournal{AJ}

\shorttitle{Outflow from Outer-arm Starburst}
\shortauthors{Kaufman et al.}

\begin{document}

\title{Outflow from Outer-Arm Starburst in a Grazing Collision Between Galaxies}

\correspondingauthor{Michele Kaufman}
\email{kaufmanrallis@icloud.com}

\author[0000-0002-5357-680X]{Michele Kaufman}
\affiliation{110 Westchester Rd., Newton, MA 02458, USA}

\author{Bruce G. Elmegreen}
\affiliation{IBM Research Division, T.J. Watson Research Center, 1101 Kitchawan Rd., Yorktown Heights, NY 10598, USA}

\author{Morten Andersen}
\affiliation{Gemini Observatory, NSF's National Optical-Infrared Astronomy Research Laboratory, Casilla 603, La Serena, Chile}

\author{Debra Meloy Elmegreen}
\affiliation{Department of Physics \& Astronomy,  Vassar College,  Poughkeepsie, NY 12604, USA}

\author{Curtis Struck}
\affiliation{Department of Physics \& Astronomy, Iowa State University, Ames, IA 50011, USA}

\author{Fr\'{e}d\'{e}ric Bournaud}
\affiliation{Laboratoire AIM-Paris-Saclay, CEA/DSM-CNRS-Universit\'{e} Paris Diderot,  Irfu/Service d'Astrophysique, CEA Saclay\\
  Orme des Merisiers, F-91191 Gif sur Yvette, France}

\author{Elias Brinks}
\affiliation{University of Hertfordshire, Centre for Astrophysics Research, College Lane,  Hatfield AL10~~9AB, United Kingdom}

\author{James C. M$^c$Garry}
\affiliation{University of Hertfordshire, Centre for Astrophysics Research, College Lane, Hatfield AL10~~9AB, United Kingdom}

\begin{abstract}
Gemini NIFS {\it K}-band spectra and ALMA \CO, HCO$^+$, and 100 GHz continuum observations are used to study
a bright starburst clump on an outer arm of  the interacting galaxy NGC 2207.  This clump
 emits  23\% of the total 24 \micron\  flux  of the galaxy pair and
has an optically-opaque dust cone extending
out of its 170 pc core.  The measured CO  accounts for the dark cone extinction if almost all the
gas and dust there is in front of  the star clusters.
 An associated approaching  CO outflow  has  $v_z \sim 16$ \kms, an
  estimated molecular mass $8 \times 10^6$ \msun, and rises to heights $\sim 0.9$ kpc.
A receding CO outflow on the far side with $v_z \sim 28$ \kms\
is less extensive. The observed star formation in the core over 10 Myr can supply the
dark cone kinetic energy of  roughly $ 2\times10^{52}$ ergs via supernovae and stellar winds.
Other signs of intense activity are  variable radio continuum, suggesting an embedded supernova or
other outburst, X-ray emission possibly from
an X-ray binary or intermediate mass black hole, depending on the extinction, and
  Br$\gamma$ and \ion{He}{1}  lines with 82 \kms\  linewidths  and fluxes
consistent with excitation by embedded O-type stars. According to previous models,
the retrograde encounter suffered by NGC 2207 caused loss of angular momentum.
 This compressed
 its outer disk. We suggest that the resulting  inward crashing stream hit
a massive \HI\ clump on the pre-existing spiral arm and triggered the observed starburst.

\end{abstract}


\section{Introduction}

The spiral galaxies \object{IC 2163} and \object{NGC 2207}
are involved in a grazing encounter \citep{elmegreen95a, elmegreen95b,struck05}.
A prominent starburst region called Feature i by \citet{elmegreen00} lies in an outer arm of
NGC 2207 on its anti-companion side. Feature i is
the most luminous 8 \micron, 24 \micron, 70 \micron, \Halpha, and radio continuum
source in NGC 2207/IC 2163 \citep{elmegreen01,elmegreen06,kaufman12} and accounts
for 23\% of the total 24 \micron\ emission from the galaxy pair \citep{elmegreen06,
elmegreen16}.
 The star formation rate (SFR) of Feature i is
1.6 M$_{\sun} {\rm yr}^{-1}$ \citep{smith14,elmegreen16}, which is enormous for a
clump not in a galactic nucleus or a galaxy merger, or in the expanding ring of a collisional ring galaxy.
The gas (H$_2$ + \HI) depletion time of Feature i, $\sim 100$ Myr \citep{elmegreen16}, lies in the
range of starburst galaxies [see, for example, the Kennicutt-Schmidt diagram in \citet{bigiel08}].

Comparable in SFR to the brightest  star-forming clump in the overlap part of the
Antennae galaxies \citep{smith14},
Feature i has the seventh highest SFR
among the set of 1700 star forming complexes in interacting galaxies or
normal spirals measured by \citet{smith16} with a 1 kpc  aperture.
Because of its high SFR and location on an outer arm, \citet{smith14} include
Feature i in their study of hinge clumps. Hinge clumps are regions of multiple converging flows,
usually  involving a tidal arm produced in a prograde encounter.  In contrast, the encounter suffered by
NGC 2207 was  retrograde and somewhat out-of-plane \citep{elmegreen95b,struck05}.
We shall consider how such an encounter may have led to the formation of Feature i.

The most striking property of Feature i is a large optically-opaque, conically-shaped dust cloud
 extending  a projected $3''$ ($\sim 500$ pc)  from the center of its $\sim 1''$ core.
This structure suggests that there is a large scale outflow of gas
 inclined with respect to the disk at Feature i and thus
feedback from the starburst.
We present new observations of Feature i in \CO, HCO$^+$, and millimeter-wave continuum
(in the range 88 GHz to 106 GHz) from
 the Atacama Large Millimeter Array (ALMA)
 and {\it K}-band spectrometry from Gemini NIFS.  We then combine these observations
with previous radio to X-ray data in order to study the outflow and determine the energy
 source(s) in this intriguing, partly obscured star-forming region.

 NGC 2207/IC 2163  contain massive ($10^8 - 10^9$ \msun)
\HI\ clouds that do not coincide with the brighter star-forming  knots in these galaxies
\citep{elmegreen93,elmegreen95a}. The massive  \HI\ clouds
 are located in the large areas of these galaxies that have high velocity
 dispersion (30 - 50 \kms) in the \HI\ gas.  Similar massive \HI\ clouds in regions with high  \HI\ velocity
dispersion are found in other interacting galaxies in an early  stage of post-encounter evolution,
 for example, Arp 82  \citep{kaufman97},
Arp 84  \citep{kaufman99}, and the NGC 5774/75 pair  \citep{irwin94}.
Presumably, the galaxy encounters
increased the \HI\ velocity dispersion and thus the gravitational Jeans mass of  the largest clouds
in these galaxies \citep{elmegreen93,wetzstein07}. Although Feature i does not presently correspond
to a massive \HI\ cloud, it is   similar to the massive \HI\ clouds in
its mass of cold gas, $\sim 2 \times 10^8$ \msun\  \citep{elmegreen16},  overall size,
$1.4 \times 1.7$ kpc \citep{kaufman12}, and location
in a region with high \HI\ velocity dispersion.
Feature i may illustrate what  a massive \HI\ cloud could become if an outburst of star formation were
triggered in it.

From the NASA/IPAC Extragalactic Database (NED), we adopt a distance of $35 \pm 2.5$ Mpc
for NGC 2207 with a Hubble constant $H$ = 73 \kms\ Mpc$^{-1}$ and corrected for
infall toward Virgo. Then $1''$ = 170 pc. Our velocities are heliocentric and use the optical
definition of the nonrelativistic Doppler shift.

 \citet{rupke10} measure [O/H] = 8.874   in Feature i, so metallicity close to solar.
We therefore adopt the standard value
$X_{\rm CO}$ = $1.8 \pm 0.3 \times 10^{20}$ H$_2$ cm$^{-2}$ (K km s$^{-1})^{-1}$ for normal
galaxies
\citep{dame01} to convert the CO emission to
molecular mass.  In Section 4 we comment on whether
a reduced value of $X_{\rm CO}$ is appropriate for Feature i.

An overview of Feature i from previous observations is presented in Section 2, and
a description of
our new data in Section 3.  Since star formation in parts of Feature i may be  hidden by heavy
obscuration, we investigate the level of extinction and the dust distribution in Section 4. The amount  of
extinction at the detected  X-ray source in  Feature i is relevant to
whether this source could be an intermediate mass black hole.  Section 5 identifies
 the two main CO components of Feature i.
In Section 6, we focus on the star-forming core of Feature i and compare its properties in the broad-band optical,
CO, HCO$^+$,
near-infrared emission lines, 100 GHz continuum and radio continuum.
We discuss the internal kinematics of the cold gas, the gas velocity dispersion, possible outflows, and the
three-dimensional orientation of the latter in Section 7. In Section 8 we hypothesize a possible origin for
Feature i, and Section 9  summarizes our conclusions.

\begin{figure*}
\epsscale{0.85}
\plotone{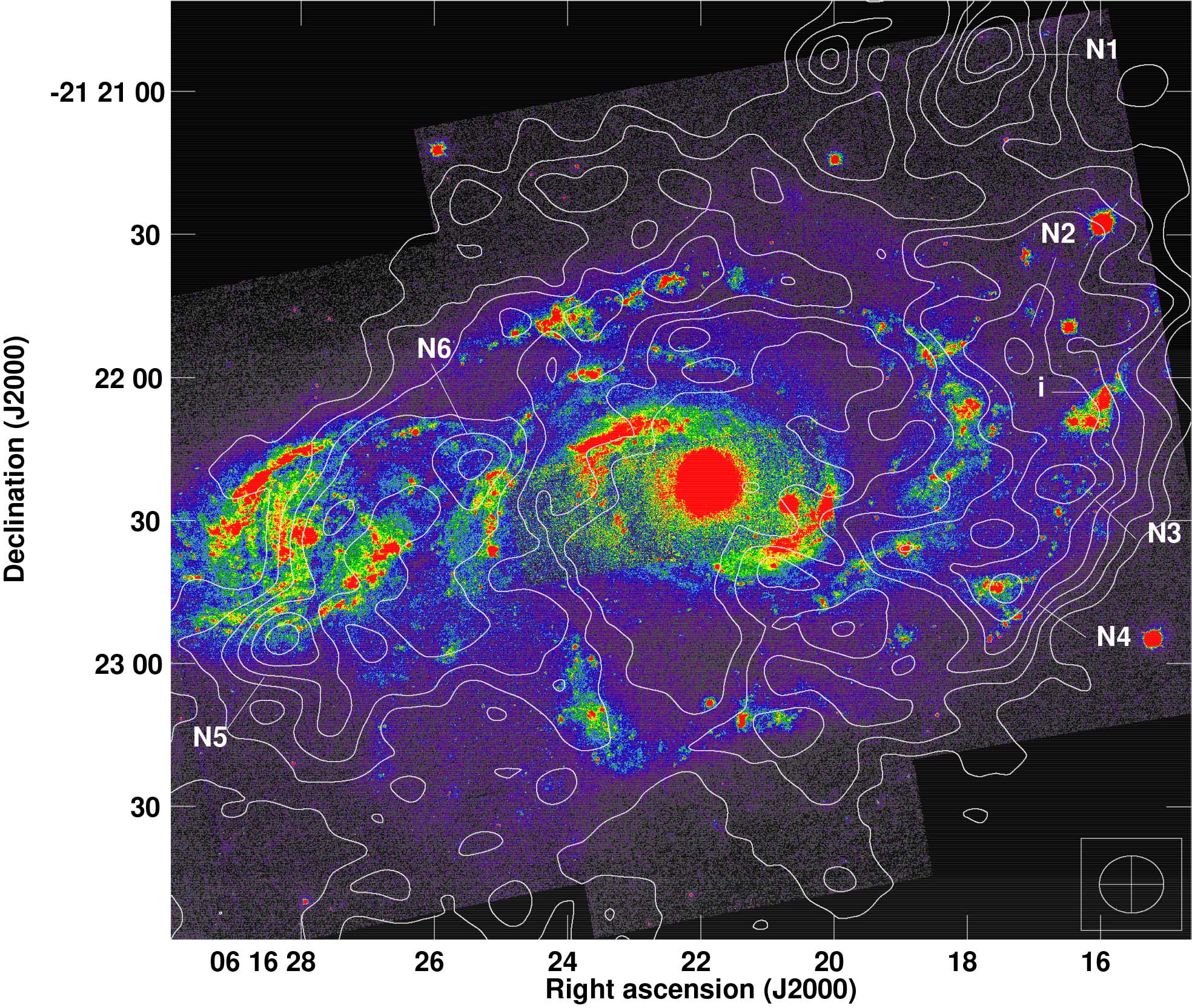}
\caption{\HI\ line-of sight column density contours of NGC 2207 overlaid on color-coded
{\it HST  B} image \citep{elmegreen00} with contours at
10, 15, 20, 25, 30, and 35 \msunpc.  Part of the companion IC 2163 is to the left.
Feature i on an outer
arm of NGC 2207 on the anti-companion side is labelled {\it i}.
The six massive \HI\ clouds in NGC 2207 are labelled N1 through N6. The beam symbol represents
the \HI\ beam.
\label{fig1}}
\end{figure*}

\begin{figure}
\epsscale{0.90}
\plotone{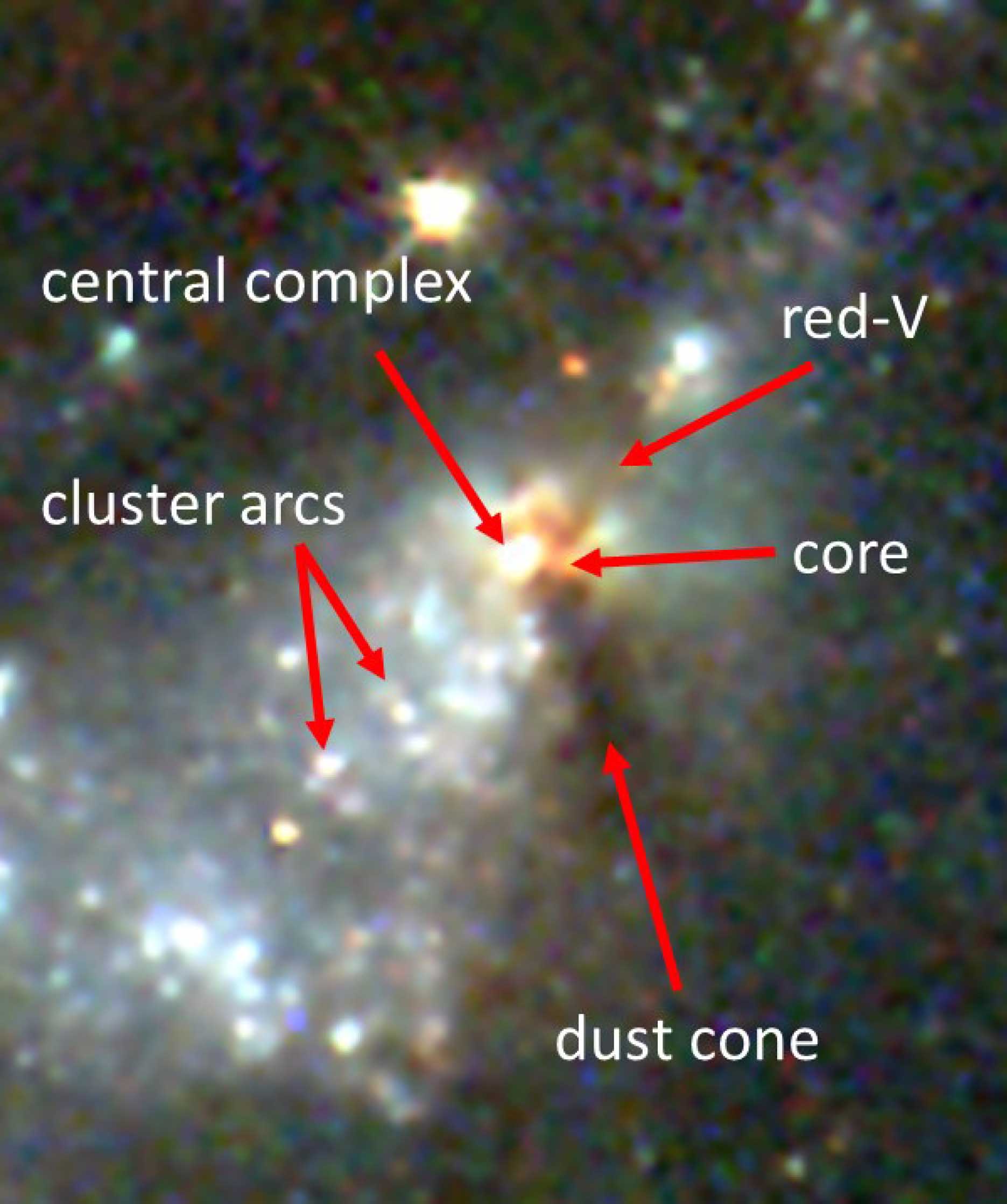}
\caption{{\it HST} color composite image of Feature i and surroundings
 with the following structures labelled:  the central star complex,
the concentric arcs of smaller clusters, the optically-opaque dust cone extending $3''$ to the
southwest of the central star complex, and
the red V-shaped dust structure opening to the north.
The field displayed is $15'' \times 18''$.
The $\sim 1''$ core of Feature i
contains the vertex of the red V, the center of the central star complex,
the apex of the dark dust cone,  and the radio continuum peak.  Within the uncertainties, the vertex
of the red V  coincides with the radio continuum peak and the {\it Spitzer} 8 \micron\ emission peak.
  The possible ULX detected with {\it Chandra} lies at, or close to, the radio continuum peak.
\label{fig2}}
\end{figure}

\section{Overview and Previous Results}

 The location of Feature i in NGC 2207 (at a distance of 15 kpc from the NGC 2207 nucleus)
 and the locations of the six massive \HI\ clouds (N1 - N6) are marked in Figure\,\ref{fig1},
which displays the {\it HST B} image overlaid with the \HI\ column density
contours of NGC 2207 \citep{elmegreen95a,elmegreen00}.   The \HI\ velocity of Feature i is
consistent with that  of the surrounding  \HI\ in the NGC 2207  disk; thus it is
likely that Feature i formed in the disk.
 Table\,\ref{Table1}  lists previous observations from radio to X-rays of Feature i, along with
the values of  the FWHM of the point spread function (PSF) and for
 the radio and mm-wave observations, the rms noise
$\sigma_{\rm eff}$ at the location of Feature i
after correcting for primary-beam attenuation.
Important components of Feature i are its core, the central star complex, the cluster arcs,
the opaque dust cone, and the red V-shaped dust structure. These are marked in
 Figure \,\ref{fig2}.
Their relevant properties are summarized below.

\subsection{Radio Continuum and 8 \micron: the Core and the Extended Emission}

In the \sixcm\ radio continuum images from NSF's
 Karl G. Jansky Very Large Array (VLA) and  in the {\it Spitzer} 8 \micron\  image,
Feature i consists of a bright core with FWHM $\sim 1''$  (= 170 pc)  embedded in a region
of extended emission
of dimension $8'' \times 9.6''$ (= $1.4 \times 1.7$ kpc); see  Figure 12 in \citet{kaufman12}.
The location of maximum surface brightness in the radio continuum is at
 R.A., decl. (J2000) = 06 16 15.865, $-21$ 22 02.77 \citep{vila90} with an uncertainty of
$0.1'' - 0.2''$ (10\% of the HPBW).  We shall use  this as the
point from which relative positions are quoted throughout this paper.

Scaled-array VLA observations show that both the core and the surrounding region are dominated by
nonthermal emission at  $\lambda 20$ cm and \sixcm.
With a $7.5'' \times 7.5''$ aperture \citet{vila90} measure flux densities $S_\nu$(20 cm) =
10.3 mJy and $S_\nu$(6 cm) = 3.4 mJy and a spectral index $\alpha = -0.9$
(where $S_\nu \propto \nu^{\alpha})$. For the $\sim 1''$ core,  their Gaussian fit gives $S_\nu$(20 cm)
= 3.4 mJy, $S_\nu$(6 cm) = 1.4 mJy, and $\alpha = -0.7$.  A near doubling of the \sixcm\
flux density  of the Feature i core between  1986 and 2001 \citep{kaufman12}
is evidence of a possible  radio supernova in the core.

\subsection{Optical: Central Star Complex, Cluster Arcs, Opaque Dust Cone, and Red V-shaped Dust Structure}

 In the {\it HST B} image,  the only star complex detected in the core of Feature i  has
a mass of $2 \times 10^6$ \msun, an age of 0.6 Myr \citep{elmegreen17}, and a size of
$0.6'' \times 0.4'' $ (100 $\times 70$ pc) with major axis at a
position angle (PA) of $-45\degr$. The center of this star complex is
$0.4''$ east, $0.4''$ south of the radio peak;  the uncertainties in position are
 $\sim 0.25''$ in the absolute astrometry of the {\it HST}
image and $0.1'' - 0.2''$ for the radio continuum peak.
 This central star complex is
surrounded by concentric arcs of 13 smaller star complexes that lie  within the radio and
8 \micron\ region of extended emission \citep{elmegreen00,elmegreen17}.
These range in age  from 0.6 Myr to 60 Myr
with an average age of 7.9 Myr and an average mass of $1.6 \times 10^5$ \msun.  There is
no obvious pattern of cluster age with distance from the radio peak in the core.

The central star complex together with a  reddish dusty region on its western edge
is near the apex of the optically-opaque dust cone
 which extends to the southwest for a projected length
of $\sim 3'' $  at  PA = $215\degr \pm 5\degr$ \citep{elmegreen00}, nearly
perpendicular to the major axis of the central star complex. The cone has an opening angle of $\sim 60\degr$;
its darkest part is a column of width $\sim 0.5''$. The reddish region also contains the vertex of a
V-shaped dust structure opening to the north. This vertex appears to coincide with the
radio continuum peak and the {\it Spitzer} 8 \micron\ emission peak.  It
 may be the location of an embedded cluster old enough to host supernovae or some other source
of nonthermal emission.

\subsection{X-rays: Possible ULX in the Core}

In four {\it Chandra} observations between 2010 and 2013, \citet{mineo14} detect Feature i
as an elongated  soft  X-ray source and suggest that it may contain a non-variable
 Ultta-luminous X-ray Source (ULX) surrounded by hot diffuse gas.
The possible ULX and the radio continuum peak coincide
within the uncertainties.

\subsection{Previous ALMA CO, \HI, and SFR}

The CO emission from Feature i  has a C-shaped distribution and consists of
 two lobes that join in the core: one lobe along the opaque dust cone southwest of the
central star complex and the other lobe along the extended  8 \micron\ emission
north-northwest of the core. The CO lobe along the dark dust cone has about the
same length  as the dark cone  \citep{elmegreen17}, and thus the extinction of the dark cone
is probably from dust associated with the cold molecular gas.

\begin{deluxetable*}{lllcl}
\tablewidth{0pt}
\tablecaption{Some Previous Observations of  Feature i\tablenotemark{a}
\label{Table1}}
\tablehead{
  \colhead{Data Set}  & \colhead{Date Observed} & \colhead{PSF (HPBW, BPA)}  &
       \colhead{noise $\sigma_{\rm eff}$}   &   \colhead{ref.} \\
& \colhead{(YYYY-mm-dd)} &  &\colhead{(\mJybeam)}    & \\
  \colhead{(1)}  & \colhead{(2)}  & \colhead{(3)}  & \colhead{(4)}  &  \colhead{(5)}   \\
 }
\startdata
Spectral Line:  &&&\\
  VLA \HI\                         & 1990-10-11  &  $13.5'' \times 12.0''$, $90\degr $   & 0.73     & 1\\
  ALMA \CO\                     & 2014-04-03   &  $2.00'' \times 1.52''$,$-68.5\degr$ & 4.4     & 2, 3\\
  Lowell \Halpha\              & 1999-10-06     & $4.2'' \times 3.6''$                    & \nodata  & 4\\
    &\\
Continuum:  &&&&\\
  VLA 1.49 GHz                & 1986-03-18  & $1.92'' \times 1.11''$, $9\degr$     & 0.19    &  5\\
  VLA 1.40 GHz                & 1990-10-11  & $10.1'' \times 6.5''$,  $90\degr$    & 0.37    &   1\\
  VLA 4.86 GHz                & 1986-08-16  & $1.93'' \times 1.04''$,$169\degr$  & 0.07    & 5\\
  VLA 4.86 GHz                & 2001-04-14 & $2.47'' \times 1.30'' $, $8\degr$    & 0.013   & 6\\
 Spitzer 24 \micron\       & 2005-03-11 & $6''$                                          &\nodata      & 7,2\\
 Spitzer  8 \micron\         & 2005-02-22 & $2.4''$                                        & \nodata   & 7\\
{\it  HST UBVI}                  & 1996-05-25 & $0.18''$                                     &\nodata     & 8,3\\
 XMM-Newton UVM2       & 2005-08-31 & $1.8''$                                       &\nodata      & 6\\
 Galex NUV                      & \nodata        & $5''$                                          &\nodata      & 9\\
 Chandra                          & 2010 -- 2013 & $1''$                                   &\nodata          & 10\\
\enddata
\tablerefs{(1) Elmegreen et al. 1995a; (2) Elmegreen et al. 2016; (3) Elmegreen et al. 2017;
    (4) Elmegreen et al. 2001; (5) Vila et al. 1990; (6) Kaufman et al. 2012;
   (7) Elmegreen et al. 2006;
   (8) Elmegreen et al. 2000; (9) Smith et al. 2014; (10) Mineo et al. 2014.}
\tablenotetext{a} {For the radio and mm-wave observations,
$\sigma_{\rm eff}$ is the rms noise at  Feature i after correcting
   for primary beam attenuation. It has units of
     \mJybeam\ per channel for the spectral line observations and \mJybeam\ for the
     continuum observations.}
\end{deluxetable*}

\begin{deluxetable*}{llllll}
\tablewidth{0pt}
\tablecaption{New ALMA Spectral Line Observations of  Feature i
\label{Table2}}
\tablehead{
  \colhead{Parameter}  & \colhead{\CO} & \colhead{HCN}  &  \colhead{HCN} &
       \colhead{HCO$^+$} & \colhead{HCO$^+$} \\
 }
\startdata
Date Observed  &  2015-05-15 &  2014-08-30  & 2015-06-07 & 2014-08-30 & 2015-06-07 \\
uv-coverage (k$\lambda$)  & $8.1 - 213$ & $8.9 - 323$ & $6.3 - 230$ & $8.9 - 323$ & $6.3 - 230$\\
Number of antennas\tablenotemark{a}  & 34  & 33    & 36           & 33                & 36\\
Time on Source  &  72 min        & 36 min            & 72 min          & 36 min          & 72 min\\
Flux Calibrator   &  Callisto       & Ganymede       & Mars             & Ganymede     & Mars \\
Central $\nu$ (GHz)  & 114.195  & 87.830        & 87.823          & 88.390          & 88.382
\enddata
\tablenotetext{a}{Omitting antennas flagged for most or all of the run}

\end{deluxetable*}

Because of the
resolution of the \HI\ data, \citet{elmegreen16} use a
$14''$ (= 2.4 kpc) diameter aperture to measure the SFR,  \HI\ mass, and
molecular mass of Feature i.
They find the  SFR = 1.6 M$_{\sun} {\rm yr}^{-1}$
(from the combination of 24 \micron\ and \Halpha\ luminosities),
the \HI\ mass $M$(\HI) = $1.1 \times 10^8$ \msun,
and  $M({\rm H}_2)$= $8.1 \times 10^7$ \msun\ (multiply
these values of the atomic and molecular mass
by a factor of 1.36 to include helium).
In this aperture, the \HI\ velocity dispersion is high (52 \kms).
The molecular gas consumption time in Feature i is only 50 Myr,
and the total gas consumption time is 118 Myr.
A somewhat smaller aperture, $10''$ in diameter,
 captures essentially all the on-going
 star formation in Feature i \citep{smith14}, the extended radio and 8 \micron\ emission,
and the molecular mass
\citep{elmegreen17}.

\section{New Observations}

\subsection{CO, HCO$^+$, HCN and Millimeter-wave Continuum}

 Table\,\ref{Table2} contains some observing details about our new ALMA \CO, HCN, and HCO$^+$
observations of Feature i, Table\,\ref{Table3} describes the resulting cubes and surface brightness
(integrated intensity) maps of the spectral lines, and Table\,\ref{Table4} lists the properties
of the line-free continuum maps associated with these observations.
The notation $\sigma_{\rm eff}$
means the rms noise at Feature i after correction for primary beam attenuation.

The ALMA \CO\ observations (rest $\nu_0$ =115.271202 GHz) on 2015 May 15
were obtained during ALMA Cycle 2.
They consist of
mosaic maps made with 34 pointings, spaced $26''$ apart, to cover the emission from the
galaxy pair, with
 phase center at R.A., decl. (J2000) =  06 16 22.809,  -21 22 30.73.
The bandwidth is  1.875 GHz.
 The brightest mm-wave continuum source in the galaxy pair is Feature i, and it  is well below
 the noise in the CO spectral-line cube.
The maximum recoverable scale is $15''$, which exceeds the size of
Feature i.

\begin{figure}
\epsscale{1.0}
\plottwo{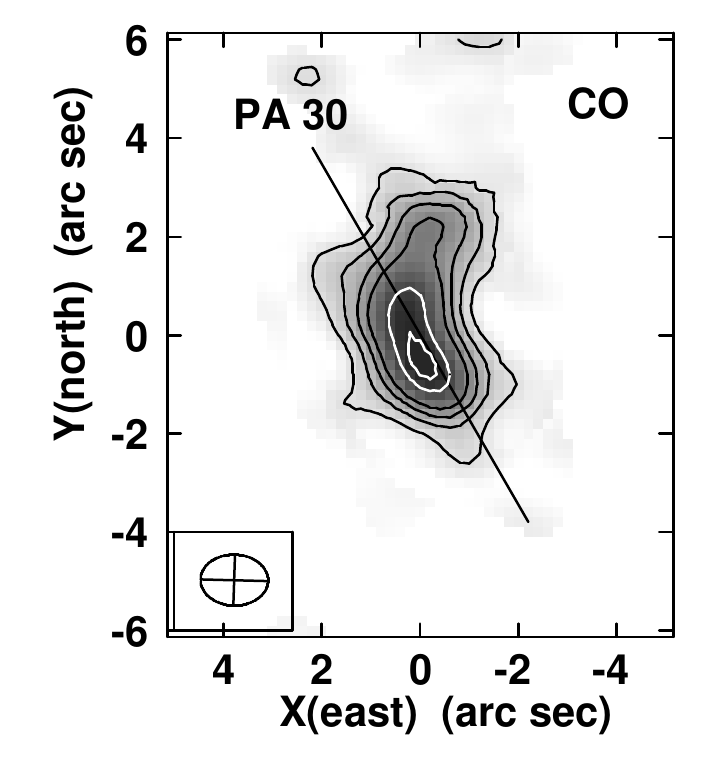}{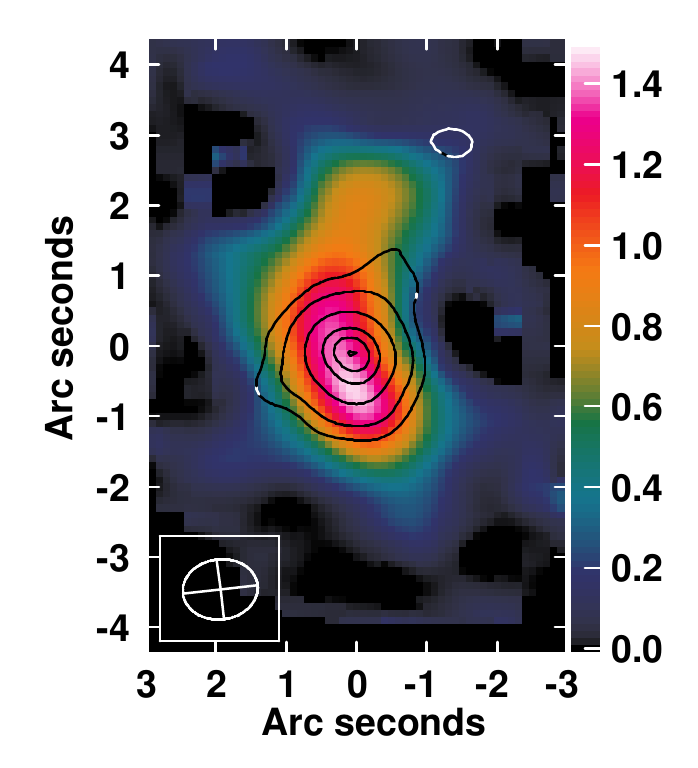}
\caption{Top: Grayscale and contour display of the CO surface brightness
with contour levels at 0.2 , 0.4, 0.6. 0.8, 1.2, and 1.4 \Jybeamkms,
where 1 \Jybeam\ is equivalent to  $T_b$ = 65.6 K and 1 \Jybeamkms\ corresponds to $N({\rm H}_2)$ = 189 \msunpc.  The axis of the dark dust cone is represented by the line drawn at
 PA = $30\degr$.
Bottom: Contours of the 100 GHz continuum emission overlaid on the color-coded  CO image
with contour levels  at
 3, 6, 12, 18, 21, 23 $\times$ the rms noise $\sigma_{\rm eff}$ of
 0.034 \mJybeam, equivalent to $T_b$ = 0.134 K.   The beam symbol represents the
100 GHz beam. The wedge is
in units of \Jybeamkms, with 1.0 units corresponding to  $N({\rm H}_2)$ = 189 \msunpc.
 The radio  and
100 GHz continuum peaks coincide, maximum CO surface brightness is displaced
$0.5''$ south of the radio peak, and the 100 GHz continuum emission is not
elongated along the dark dust cone.  Both panels are centered on the radio peak.
\label{fig4}}
\end{figure}

 Using the pipeline with CASA 4.2.2 software, the ALMA Data Reduction Team did the calibration and flagging of the \CO\ observations and made and cleaned a data cube of the
CO Cycle 2 line emission.
Inspecting this, we
saw no problems. For subsequent analysis, we used the AIPS software
package. To select areas of genuine emission in the CO data,
we convolved the cube before correction for
primary-beam attenuation to $3''$ resolution, clipped it at 2.5 times its rms noise , and
retained regions of emission only if  they appear in at least 2 adjacent velocity channels.
The result was applied as a blanking mask to the original cube, and after correcting for
primary beam attenuation, we calculated the moment maps. Additionally, we blanked the
 intensity-weighted velocity field (first moment) and the velocity dispersion
(second moment) images where the CO surface brightness is less than
($2.5 \times$  rms noise) $\times $(2 channel widths).
Figure\,\ref{fig4}
displays the CO surface brightness (integrated intensity) map of Feature i from this cube.

\begin{deluxetable*}{lllll}
\tablewidth{0pt}
\tablecaption{ New ALMA Spectral Line Cubes and Maps of Feature i
\label{Table3}}
\tablehead{
    \colhead{Parameter}  & \colhead{CO}  & \colhead{CO}  & \colhead{HCN}  &  \colhead{HCO$^+$}  \\
                    &    \colhead{Cycle 2}  &  \colhead{Combined} \\
}
\startdata
Channel width      & 10 \kms\   & 4.87 \kms\     &  8 \kms\        & 10 \kms\  \\
Weighting            & Briggs R = 0.5  & Briggs  R = 0.5  & Natural   &   Natural    \\
PSF (HPBW)   & $1.37'' \times 1.04''$  &  $1.82'' \times 1.33''$   &
                         $1.22'' \times 0.95''$   &  $1.26'' \times 1.00''$ \\
PSF (BPA)             & $88.8\degr$   & $-73.4\degr$   & $-85.6\degr$   & $-81.7\degr$ \\
Pixel size             & $0.2''$  &   $0.2''$    &  $0.3''$   & $0.3''$  \\
 $T_b/I$\tablenotemark{a} (K/\Jybeam)  & 65.6   & 38.6     &  137        & 124   \\
At Feature i: &&&\\
 $\sigma_{\rm eff}$ per channel\tablenotemark{a}
                            &  3.8     & 3.3       &  0.96                   & 0.86  \\
peak $S/N$\tablenotemark{b}   &  12   & 17   &    $< 2$      &  3.7  \\
Max. brightness\tablenotemark{c}  &  1.45  &  2.44 &  \nodata   & 0.062\\
&& \\
$N({\rm H}_2)$ equivalent  to\\
 1 \Jybeamkms\   in CO  & 189 \msunpc\     & 111 \msunpc\ & \nodata   & \nodata \\
max. $N({\rm H}_2)$ in Feature i  & 281 \msunpc\  & 271 \msunpc\  & \nodata  & \nodata
\enddata
\tablenotetext{a} {$T_b/I$ gives the  brightness temperature
$T_b$ equivalent to a surface brightness  $I$ of 1 \Jybeam, and
$\sigma_{\rm eff}$ (in \mJybeam) is the rms noise at
Feature i after correcting  for primary beam attenuation.}
\tablenotetext{b} {Peak $S/N$ in channel maps}
\tablenotetext{c} {Maximum surface brightness (\Jybeamkms) in integrated intensity image
of Feature i}
\end{deluxetable*}

\begin{deluxetable*}{llll}
\tablewidth{0pt}
\tablecaption{ALMA  Continuum Maps of Feature i
\label{Table4}}
\tablehead{
  \colhead{Parameter}   & \colhead{88 GHz\tablenotemark{a}}  & \colhead{100 GHz\tablenotemark{a}}
    & \colhead{106.4 GHz\tablenotemark{b}}\\
}
\startdata
Bandwidth         & 0.52 GHz  & 3.9 GHz          & 6 GHz\\
Weighting          &  Natural    & Natural            &  Briggs R= 0.5\\
PSF (HPBW)        & $1.26'' \times 1.00''$    & $1.07'' \times 0.85''$
                           &  $1.64'' \times 1.22''$ \\
PSF (BPA)           & $- 81.7\degr$       & $-83.5\degr$    &  $-79.7\degr$ \\
Pixel size          & $0.1''$      & $0.1''$             &  $0.2''$  \\
$T_b/I$  (K/\Jybeam)  & 125 & 134                 &  54.3\\
 $\sigma_{\rm eff}$\tablenotemark{c}  (\mJybeam) & 0.068 & 0.034  & 0.044 \\
 Max. Surface Brightness\tablenotemark{d}  & 1.33 & 0.79  &  1.04
\enddata
\tablenotetext{a} {made along with HCN and HCO$^+$ observations on 2014-08-30 and 2015-06-07}
\tablenotetext{b} {made along with CO Cycle 2 observations on 2015-05-15}
\tablenotetext{c} {$\sigma_{\rm eff}$ is  the rms noise at Feature i after applying
  the correction  factor for primary beam attenuation.}
\tablenotetext{d} {Units are  \mJybeam}
\end{deluxetable*}

We also combined our lower resolution ALMA Cycle 1 CO data
(listed in  Table\,\ref{Table1})
with our Cycle 2 CO data, then made and cleaned a data cube of the combined emission,
and similarly masked it.  The cube derived from the Cycle 2 data alone is called the
CO Cycle 2 cube, and the one derived from the combined Cycle 1 plus Cycle 2 data is
called the CO Combined cube.   The CO Cycle 2 cube  has  the higher
spatial resolution. The CO Combined cube has the higher velocity resolution and
is more sensitive to extended emission. For the $\sim 1''$ core and for where a clearer distinction between the core and the dark cone is needed, we use the CO Cycle 2 cube.
For the velocity field in Section 7, we use the CO  Combined Cube.  Unless
otherwise noted, our results are from the CO Cycle 2 cube, although for some applications,
such as the extinction on a kpc scale (Section 4.1), both CO cubes yield the same values.

On 2014 Aug 30  and on 2015 June 7,  we made ALMA observations of the spectral lines
HCN (1--0)
(rest $\nu_0$ = 88.631601 GHz) and HCO$^+$ (1--0)  (rest $\nu_0$ = 89.188526 GHz)
and also observations of the 100 GHz continuum.
On 2014 Aug. 30, the maximum recoverable scale is $11'' - 12.5''$ for observations in the
range 100 - 88  GHz; on 2015 June 7, it is $17'' - 20''$ for these frequencies.
Three fields in the galaxy pair were observed.
The westernmost field has phase center RA, decl. (J2000) = 06 16 17.654, - 21 22 13.36
and includes Feature i near its western edge.
Using the Pipeline-Cycle 2-R1-B and CASA 4.2.2 software,
the ALMA Data Reduction Team did the calibration and flagging, combined the uv-data from the two
dates, and made and cleaned the continuum image and data cubes of the line emission.

In addition to the 100 GHz continuum map,
we made a continuum map at 88 GHz from the line-free channels
in the HCO$^+$ cube.
 In the  HCO$^+$ channel maps,
the maximum S/N is only  $3.7 \times \sigma_{\rm eff}$.
To make a blanking mask for the  HCO$^+$ cube, we followed the same procedure as
with the CO data except that we convolved the HCO$^+$ cube to $2''$ resolution
rather than $3''$ resolution and  used clip level of  2 times its rms noise.
In the HCO$^+$ integrated intensity image
 at Feature i, a surface brightness $I({\rm HCO}^+)$ of (2 $\sigma_{\rm eff}) \times$ (2 channel widths)
= 0.0344 \Jybeamkms, and the maximum value of  the surface brightness
$I({\rm HCO}^+)$  is 0.062 \Jybeamkms, displaced $0.4''$ north, $0.5''$ east of the radio continuum peak.

No  HCN emission was detected in Feature i. This provides an upper limit on its HCN/CO ratio
(see Section 6).

\begin{figure*}
\epsscale{0.75}
\plotone{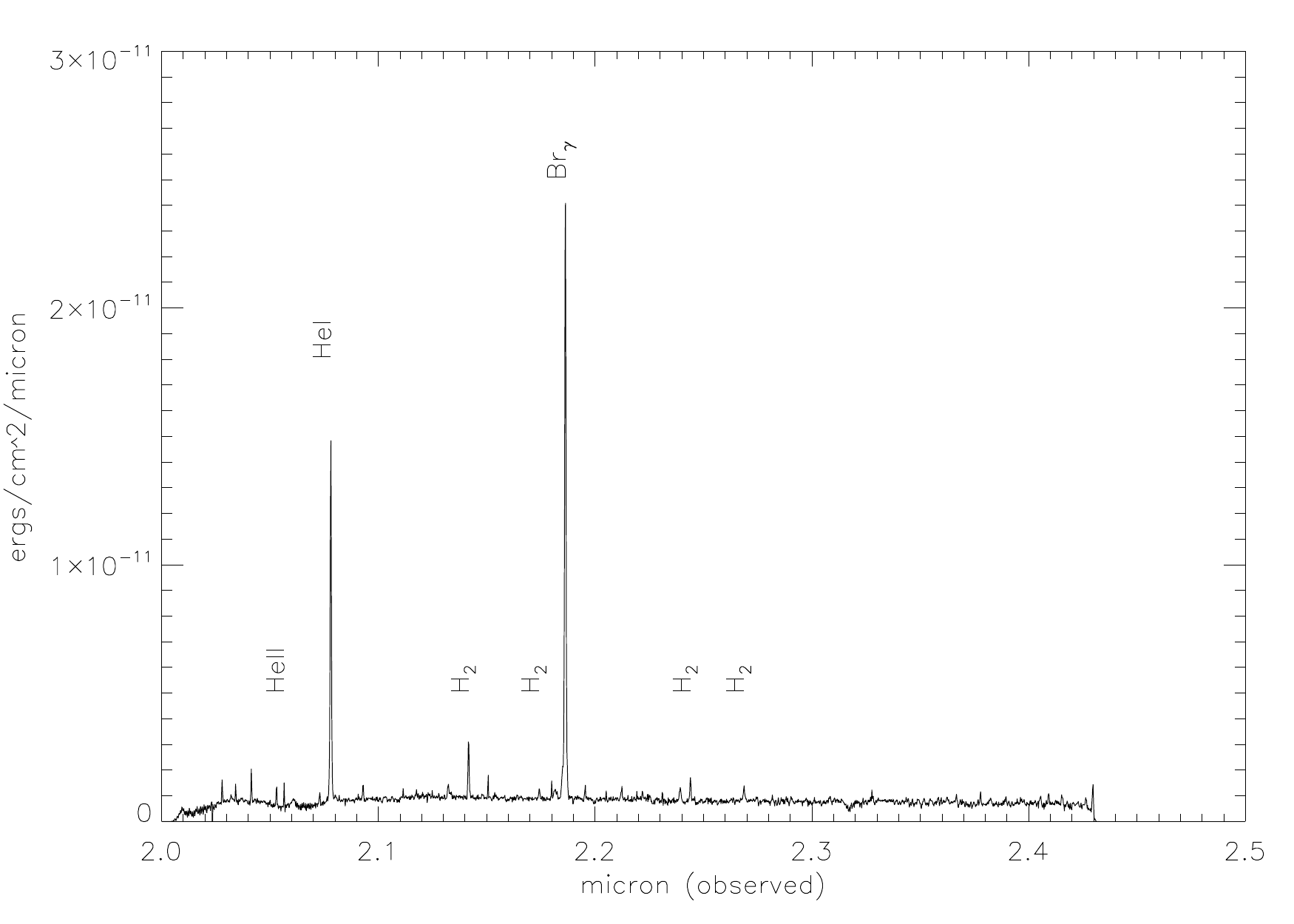}
\caption{{\it K}-band spectrum of the core of Feature i from photometry with an aperture $1.1''$ .
The abscissa is the observed wavelength. Emission lines of  Br$\gamma$, \ion{He}{1},  \ion{He}{2},
and the ro-vibrational lines of H$_2$ are marked.
\label{irspectrum}}
\end{figure*}

\subsection{Near-IR {\it K}-band Spectra}

On 2017 Feb 15, {\it K}-band spectra were taken of the core of Feature i with Gemini's NIFS
integral field spectrograph under photometric conditions.
NIFS is an image slicer Integral Field Unit (IFU) with a field-of-view  3\arcsec\
and a pixel scale of $0.1'' \times 0.04''$.
 Due to the size of Feature i and the lack of suitable guide stars in the $3''$ field,
 we observed in natural seeing mode.
The resulting spatial resolution is estimated to be  $\sim 0.5''$ or better. The precise
R. A., decl. at field center is less well determined than if there were a guide star within
the $3''$ NIFS field.

Observations were done in a traditional ABBA pattern, to remove the sky contribution, repeated 9 times with 150 seconds exposures, where each exposiure is an average of two coadds.
The ``A'' position was centered on Feature i and the ``B'' position on a piece of blank sky  identified in previous imaging.

The science data were reduced in a standard manner within the pyraf environment using the Gemini nifs package.
Darks were subtracted and clean running sky frames constructed from the ``B''  frames.
Wavelength calibration was performed using an Argon/Xenon lamp; the uncertainty in this calibration
is less than 0.2 pixels in wavelength, equivalent to  12 \kms.
Since the spatial extent of  the Br$\gamma$ emission determined by fitting a 2-D Gaussian is
 $1.1''$ (= 190 pc) (FWHM), we used a 1.1\arcsec\ diameter aperture for extracting the
spectrum of Feature i from the combined data cube.

A telluric star was observed before (HIP 23671) and after (HIP 34949) the science sequence in order to be able to correct for the absorption features in the spectra due to the Earth's atmosphere.
These were further used to flux calibrate the spectrum of Feature i, with a 1.1$\arcsec$
diameter aperture. Figure\,\ref{irspectrum}  presents the final corrected, calibrated,
background-subtracted  {\it K}-band spectrum of  Feature i obtained with the  $1.1''$ diameter aperture.
We identified redshifted Br$\gamma$, H$_2$,  \ion{He}{1}, and  \ion{He}{2}
 lines in the spectrum and performed a Gaussian fit to each line to measure its central wavelength,
line-width, and flux.
The instrumental line-width estimated from the width of the arc lines is $1.6 \times 10^{-4}$ \micron, equivalent to 22 \kms\ at  Br$\gamma$.
This was then subtracted in quadrature from the line-widths of the spectral lines detected in Feature i.

\begin{deluxetable}{lllc}
 \tablewidth{0pt}
 \tablecaption{Mean Extinction in Various Parts of Feature i
 \label{TableAv}}
\tablehead{
   \colhead{Location} & \colhead{Indicator} & \colhead{$A_{\rm v}$\tablenotemark{a}} & \colhead{ref.} \\
     &              &                   \colhead{(mag)}  \\
 }
 \startdata
 $7.2''$ aperture\tablenotemark{b}  &  H$\alpha / \lambda 6$ cm  &   $< 4.9 $  &  1  \\
 $7.2''$ aperture\tablenotemark{b}   &  H$\alpha / \lambda 6$ cm &   $\geq 3.2 \pm 0.4$  & 2 \\
 $7.2''$ aperture\tablenotemark{b}  &  $L(24 \micron)$/$L({\rm H}\alpha)$  & 3.5 & 3, 2  \\
 $7.2''$ aperture\tablenotemark{b}   & $N$(H$_2$)                     &      $2.2 \pm 0.2$  & 2  \\
 $7.2''$ aperture\tablenotemark{b}  &  $N$(H$_2$)  + $N$(\HI)  &      $\sim 3$    & 2       \\
  central star complex                      &  {\it HST  UBVI}                &       3.3   &  4 \\
 13 star complexes\tablenotemark{c}  &  {\it HST UBVI}                &  $1.45 \pm 0.76$  & 4 \\
brightest H$\alpha$ emission\tablenotemark{d}   & H$\beta$/H$\alpha$ &  0.74  & 5\\
$1.6''$ aperture on core\tablenotemark{b} & $N$(H$_2$)   & 7.5  & 2\\
At radio peak                                    & $N$(H$_2$)   & 8.3  & 2 \\
dark cone outside core\tablenotemark{e} & $N$(H$_2$)   & 4.7\tablenotemark{f}  & 2
 \enddata
\tablerefs{(1) Kaufman et al. 2012; (2) this paper; (3) Smith et al. 2014; (4) Elmegreen et al. 2017;
  (5) Rupke et al. 2010}
\tablenotetext{a} {$A_{\rm v}$ to midplane unless otherwise noted}
\tablenotetext{b} {centered on radio peak}
\tablenotetext{c} {mean of star complexes in cluster arcs outside of core}
\tablenotetext{d} {$1''$ diameter aperture centered on brightest H$\alpha$ emission, which is
$0.85''$ east, $2.3''$ south of radio peak and thus outside of the core.  The  H$\alpha$ position
uncertainty is $\sim 1''$.}
\tablenotetext{e} {$1.6''$ diameter aperture centered at distance $d = 2''$ from radio peak along
   PA = $217\degr$}
\tablenotetext{f} {$A_{\rm v}$ is front-to-back for dark cone outside core.}
\end{deluxetable}

\section{Extinction in Feature i}

 Table\,\ref{TableAv} lists estimates of  $A_{\rm v}$ in various parts of Feature i.
Except in the case of the ``dark cone outside core,'' these are
$A_{\rm v}$ values to the midplane under the assumption that the gas and dust are
symmetrically distributed about the  midplane. Some details, discussion,  and  the molecular
column densities are given below.

\subsection{Extinction on kpc Scale}

For the extinction  in a $7.2''$ (1.2 kpc) diameter aperture, which
contains most of the CO emission from Feature i, all three types of methods in
Table\,\ref{TableAv} are consistent with $A_{\rm v}$= 3 - 3.5 mag to the midplane.
From the   H$\alpha / \lambda 6$ cm radio continuum ratio, the upper limit to
$A_{\rm v}$ is obtained by taking the radio continuum emission as entirely thermal
(which it is not), and the lower limit assumes the thermal fraction of the $\lambda$20 cm
radio continuum emission is $(10 \pm 3$)\%.
From the $L(24 \micron)$/$L({\rm H}\alpha)$ ratio,  \citet{smith14} find $A_{\rm v}$ equals 4 mag
with a $10''$ diameter aperture. Adjusting to the $7.2''$ aperture gives 3.5 mag.

The mean line-of-sight column density $N$(H$_2$)  of CO
emission in the $7.2''$ aperture is  $(4.1 \pm 0.4) \times 10^{21} $ molecules cm$^{-2}$ (=
$66 \pm 7$ M$_\odot$ pc$^{-2}$).
 Our \HI\ observations, which did not resolve Feature i,  have a mean atomic
column density $N$(HI) = $3.4 \times 10^{21}$ atoms cm$^{-2}$ in the
$13.5'' \times 12''$  \HI\ HPBW.   As we do not know about  the clumping of \HI\
within the \HI\ synthesized beam, we adopt this mean value for $N$(HI) here.
With the standard Galactic relation between
neutral hydrogen column density and $A_{\rm v}$ \citep{bohlin78},
the resulting $A_{\rm v}$  from front to back is $4.3 \pm 0.4$ mag from the
molecular hydrogen and 1.8 mag from \HI,  for a total front-to-back
$A_{\rm v}$ of  6.1 mag  and thus $\sim 3$ mag to the midplane if the gas and dust are
 symmetrically distributed about the midplane.

\subsection{Extinction in Core and in Dark Dust Cone}

The gas column density gives a much higher value of the extinction in the core of  Feature i
than the value deduced from the {\it UBVI} magnitudes of the 0.6 Myr central star complex
(see Table\,\ref{TableAv}).
With a $1.6''$ diameter aperture (to approximate the synthesized beam area of the
CO Cycle 2 data)
centered on the radio continuum peak,
the mean value of
$N(\rm {H}_2$)   is  $(1.41 \pm 0.04) \times 10^{22}$ molecules cm$^{-2}$
(= $226 \pm 7$ M$_\odot$ pc$^{-2}$), equivalent to a front-to-back  $A_{\rm v}$ of  15 mag
based on the molecular gas alone.
It is dominated by the portion of the opaque
dust cone in this aperture. Maximum CO surface brightness, located at
 the apex of the dark cone,  is displaced  $0.5'' \pm 0.17''$
south of the radio continuum peak (see Figure\,\ref{fig4}).
The molecular column density
$N(\rm {H}_2$)   is  $1.75  \times 10^{22}$ molecules cm$^{-2}$ at the CO surface brightness
maximum and  $1.62 \times 10^{22}$ molecules cm$^{-2}$ at the radio peak.
The latter is equivalent to a front-to-back A$_{\rm v}$ of 16.5 mag based on the molecular gas alone
 plus a contribution (possibly 1.8 mag as in Section 4.1) from   $N$(\HI).
 If the gas at the radio peak were
distributed symmetrically about the midplane, then  $A_{\rm v}$ = 8.3 mag to the midplane
from dust associated with the molecular gas there.
Although starbursts tend to have lower values of  $X_{\rm CO}$ by a
factor of  $\sim 4$,  with a large dispersion in the amount of reduction   \citep{bolatto13},
the extinction required to account for the  opaque dust cone  implies that the value
of $X_{CO}$ on the dark cone  (where the extinction is too high to be measured optically)
is not appreciably reduced from the standard value.

Measuring  $N(\rm {H}_2$)  farther out along the optically opaque  dust cloud, specifically
with a $1.6''$ diameter aperture centered at distance $d = 2''$ from radio peak
along  PA = $217\degr$, gives  $N({\rm H}_2$)  =
$(4.4\pm 0.4) \times 10^{21}$ molecules cm$^{-2}$ (= $70 \pm 7$ \msunpc),
equivalent to a front-to-back extinction of 4.7 mag from molecular gas alone.
If this gas were symmetrically distributed about the midplane, then A$_v$ to the midplane
would be 2.3 mag from molecular gas plus a contribution from \HI.   If we use the same value
of $N$(\HI) as in Section 4.1, then $A_v$ to the midplane = (2.3 + 0.9) mag
= 3.2 mag, not sufficient to explain the dark dust cone which is optically thick even in I-band.
The implication is that almost all of the gas in this part of the dark dust cone must be on
the near side (relative to us) of the NGC 2207 disk.

\subsection{Comment about Extinction and X-rays from Feature i}

  The possible ULX detected with {\it Chandra} lies at, or close to,
the radio continuum peak, and is
slightly offset to the northwest of the  0.6 Myr central star complex. This star complex  is
too young to host a ULX.
Neither \citet{mineo14} nor \citet{smith14} measure the X-ray extinction $A_X$ within the
Feature i core.
Assuming an $A_v$ of 4 mag to the midplane, \citet{smith14} deduce the value of
$A_X$ and get a de-reddened
$L_X \sim 2 \times 10^{40}$ erg s$^{-1}$ for the 0.5 - 10 keV luminosity of the
central source in Feature i.  Their value of $L_X$ can be accommodated by a stellar mass black hole
with geometric beaming \citep{madau88} or inhomogeneous accretion \citep{ruszkowski03}.
Since the extinction at the potential ULX is probably much higher than \cite{smith14} assumed
 (see Table\,\ref{TableAv} regarding $A_v$ at the radio peak),
 the de-reddened value of $L_X$ may be considerably greater than this,
and thus the possible ULX could be an intermediate mass black hole.

\begin{figure}
\epsscale{0.88}
\plotone{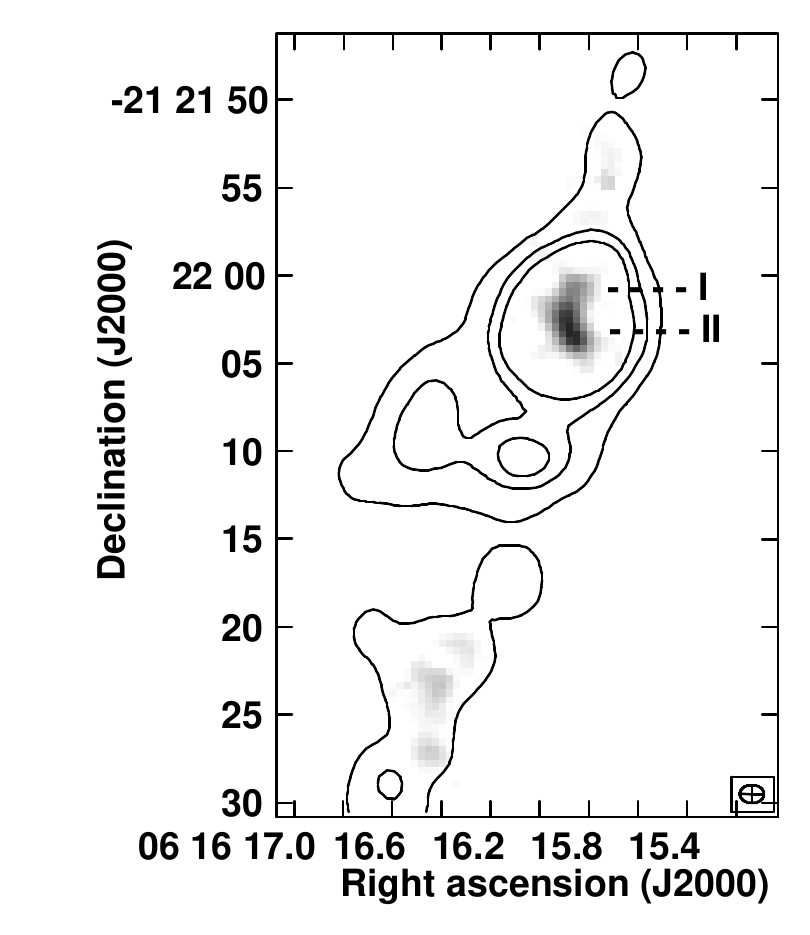}
\caption{ Location of the spiral arm near  Feature i is outlined by  contours of
{\it Spitzer} 8 \micron\ emission at
4, 8, and 12 MJy sr$^{-1}$. These are overlaid on  $I$(CO) in greyscale.
 The CO lobe of Feature i along the extended  8 \micron\ emission NNW of the core is labelled I, and the CO lobe along
the dark dust cone is labelled II.
\label{arm}}
\end{figure}

\section{The Two Lobes of  CO Emission}

 In Figure\,\ref{arm} the location of the spiral arm near  Feature i is outlined by a few
contours of 8 \micron\ emission overlaid on CO in greyscale.
CO lobe I, marked in this figure, lies along the major axis of the
 extended 8 \micron\ emission of Feature i north-northwest of the core
at PA = $-25\degr$, and thus is molecular gas in the disk of NGC 2207.
In the core, it joins CO lobe II, which coincides with the opaque dust cone at a PA of $210\degr$.
As pointed out in Section 4.2, lobe II needs to be in front of the disk to account for
the extinction of the dark cone.

According to  encounter simulation models \citep{elmegreen95b,struck05},
the minor axis of the projection of the main disk of NGC 2207 into the sky plane
is at PA = $50\degr  - 70\degr$, with the near side (relative to us) in the northeast,
 and the central disk of NGC 2207 has an inclination $i$= $35\degr \pm 5\degr$
 ($i$ = $0\degr$  for face-on).
The opaque dust cone  extends towards the southwest of the core at PA =
$215\degr \pm 5\degr$, not far from the  minor axis of this projection.
Thus, unless the disk at Feature i is highly warped, gas flowing towards us from the
center of the Feature i core in a direction quasi-perpendicular to the NGC 2207 disk
 would appear in projection to lie along the  dark dust  cone.

Figure\,\ref{fig4}  reveals that some of the  CO emission
 with $I$(CO) $\geq 1.2$  \Jybeamkms\ extends northeast
of the radio peak for $\sim 1''$ along a PA not far from that of the dust cone on the opposite side.
 Compared to the opaque cone, this bright northeast CO emission region has less extinction
(see  Figure \,\ref{fig2}) and thus less gas and dust in front of the midplane.
Section 7 presents kinematical evidence that the dark cone is from gas approaching us
 on the near side of the midplane and some of the bright northeast CO emission may
be from gas receding from us on the far side.

\section{Core of Feature i and Left Fork of Red V}

\subsection{Overview of CO, 100 GHz Continuum, HCO$^+$, and HCN}

 The resolution of the CO, 100 GHz continuum,  and HCO$^+$ images
is about the same as the FWHM of  the core (see Table\,\ref{Table3}  and Table\,\ref{Table4} ).
The bottom panel of Figure\,\ref{fig4} displays contours of the 100 GHz continuum emission
of Feature i overlaid on the  CO surface brightness image.
The 100 GHz continuum emission is not elongated along the dark dust cone.
 The locations of
maximum surface brightness in the 100 GHz continuum,
 radio continuum, and  {\it Spitzer} 8 \micron\ emission coincide
within the uncertainties.

\begin{figure*}
\epsscale{0.67}
\plotone{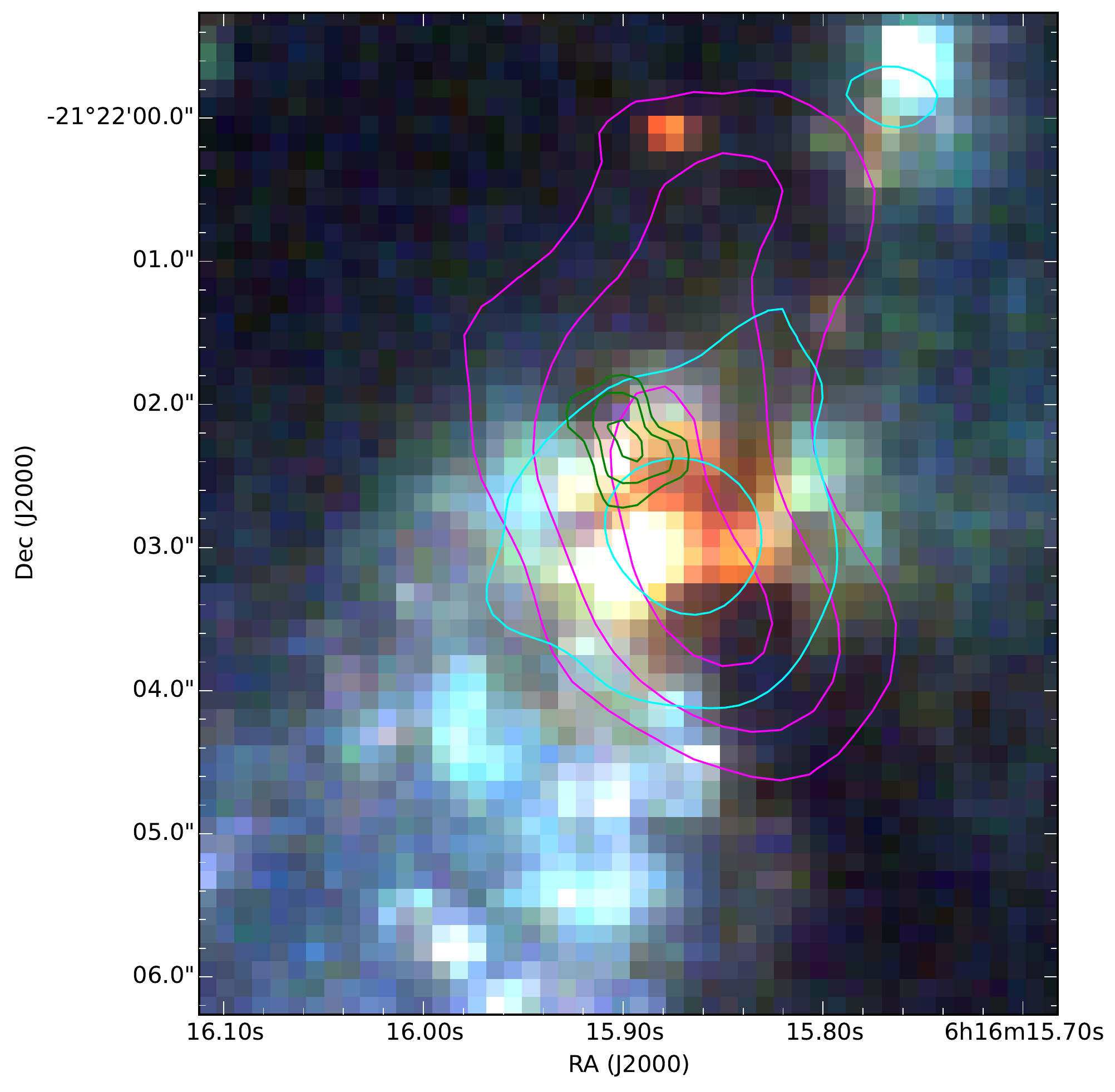}
\caption{A few indicative surface brightness contours with CO in magenta, the 100 GHz
continuum in cyan, and  HCO$^+$ in green  overlaid on an {\it HST} color composite image
of Feature i.  This shows the correspondence between the dark dust cone and the southwest lobe of CO emission.   Figure\,\ref{fig4} displays additional intermediate-level contours
of CO and the 100 GHz continuum.
The contour levels of HCO$^+$ are at 0.040, 0.050, and 0.060 \Jybeamkms,
where 1  \Jybeam\ is equivalent to
$T_b$ =  124 K, and (2 $\sigma_{\rm eff}) \times$ (2 channel widths) = 0.0344 \Jybeamkms.
 The HCO$^+$ emission
is diametrically opposite the CO emission from the dark cone and appears to lie roughly along the left
fork of the red V-shaped dust structure opening to the north as seen in the {\it HST} image.
 The spatial resolutions are $\sim 1.2''$ for the CO and HCO$^+$ images,
and $\sim 1.0''$ for the 100 GHz continuum.
\label{colorcontours}}
\end{figure*}

Figure\,\ref{colorcontours}, which overlays a few representative contours of  HCO$^+$, CO, and 100 GHz emission in
different colors on the {\it HST} color composite image of Feature i,  reveals that the
HCO$^+$ emission lies  roughly along the left fork
of the red V in the {\it HST} image and diametrically opposite the CO emission from
the dark dust cone.
In the HCO$^+$ image,  emission at the 0.035 \Jybeamkms\ level
extends along  PA $\sim 40\degr$ from a radial distance  $r =0.3''$ to $r = 1.3''$
relative to the location of radio continuum peak.
  The location and elongated shape of
the HCO$^+$ emission suggest that it may be produced by shock  heating and
excitation in a collimated wind from the energetic source at the
radio  peak (see the velocities in Section 7.1).

 The critical density of  HCO$^+$ (1 - 0) is 100 times that of CO (1 - 0)
\citep{juneau09}.
 Figure\,\ref{colorcontours}  (see Figure\,\ref{fig4} for additional
intermediate-level contours
of CO and the 100 GHz continuum) indicates that the dense gas traced by
HCO$^+$ is more compactly distributed than the lower density molecular gas detected
in CO or the 100 GHz continuum emission.  The
HCO$^+$ image has a lower S/N than the CO or the 100 GHz
continuum images.   Also the CO distribution is somewhat smoother and more widespread
than the HCO$^+$  probably because the dense structures seen
in HCO$^+$  are only a small fraction compared to the  CO-emitting gas;
the  HCO$^+$ may trace a few small clumps inside of larger molecular clouds.

The goal of  HCN observations was to search for even denser molecular gas.
No  HCN emission is detected in Feature i.  In particular we do not detect HCN emission
along the left fork of the red V where  HCO$^+$ is detected, even though both have
about the same value of   $\sigma_{\rm eff}$.
As an HCN upper limit for the core, we take $I$(HCN) =
(2 $\sigma_{\rm eff}) \times$ (2 channel widths) = 4.2 K \kms.
For CO, we use the mean surface brightness $I$(CO)= 78.7 K \kms\
in a $1.6''$-diameter aperture centered on the radio peak, as this approximates the CO synthesized
beam area. This gives an upper limit to the surface brightness ratio $I$(HCN)/mean $I$(CO) of  0.05.
 Using the same aperture for HCO$^+$,
we obtain a mean $I({\rm HCO}^+)$ = 3.4 K \kms\ and
thus the ratio $I({\rm HCO}^+)$/$I$(CO) = 0.044 in the Feature i core.
Some of the measured CO emission of the core is  from  molecular
gas ejected quasi-perpendicular to the stellar disk (see Section 7).

 Feature i as a whole has a lower  H$_2$/\HI\ ratio  than  almost all of the positions in
the studies by \citet{usero15} and \citet{bigiel16} of HCN-to-CO
and FIR-to-HCN ratios at various locations in other disk galaxies. These authors note that
clumps with a low H$_2$/\HI\ ratio tend to have a lower HCN-to-CO ratio;
the upper limit to HCN/CO in the Feature i core seems consistent with this trend.
Comparable in SFR to
the clump with the highest SFR in the overlap part of the Antennae galaxies, Feature i
may be similar to it in terms of tracers of dense gas.  For this Antenna clump
 (which contains SSC B1),
 \citet{bigiel15} measure a HCN-to-CO luminosity ratio of 0.027
 (which is below our  upper limit for the Feature i core), and,
 at the location of its HCN peak,
an $I({\rm HCO}^+)$/$I$(CO) ratio of 0.055  (25\% higher than what we find in the Feature i core),
 and an $I({\rm HCO}^+)$/$I$(HCN) ratio of 1.67.

 In comparison, the M82 central starburst, because of its high
$I({\rm HCO}^+)$/$I$(HCN) ratio of 1.50 \citep{krips08} and numerous SNe,
is considered  a late-stage starburst in which  HCN is being  depleted,
either consumed in star formation or dissipated by turbulence  or expelled via winds and feedback
\citep{baan08,krips08,kepley14}.  In less dense gas, HCO$^+$ is more efficiently excited
than HCN \citep{costagliola11}. Feature i  may also be a late-stage starburst.
For Feature i as a whole, \citet{elmegreen16}
note that the molecular gas consumption time is 50 Myr.

\begin{deluxetable*}{lccccc}
\tablewidth{0pt}
\tablecaption{Emission Lines in {\it K}-band Spectrum of Feature i Core\tablenotemark{a}
\label{IRTable}}
\tablehead{
  \colhead{Line}  &  \colhead{$\lambda_0$} & \colhead{$\lambda_{\rm obs}$}  &
    \colhead{$v$\tablenotemark{b,c} } &
    \colhead{$\sigma_v$\tablenotemark{b,d} } &  \colhead{$F$\tablenotemark{b,e}} \\
    & \colhead{(\micron)}  & \colhead{(\micron)}  &  \colhead{(\kms)} & \colhead{(\kms)}   &
}
\startdata
Br$\gamma$         &  2.16612    & 2.18625    & $2786 \pm 1$  & $35 \pm 1$  & $19.7  \pm 0.1$\\
H$_2$ 1 - 0 S(2)\tablenotemark{f}  & 2.03376    & 2.05302    &  $2839 \pm 3$  & $12 \pm 3$  & $0.42\pm 0.02$\\
H$_2$ 1 - 0 S(1)    &  2.12183    & 2.14160   & $2793 \pm 1$  & $27 \pm 1$  & $1.49  \pm 0.03$\\
H$_2$ 1 - 0 S(0)   &  2.22329    & 2.24394    & $2784 \pm 3$  & $25\pm 3$  & $ 0.77 \pm 0.02$  \\
H$_2$ 2 - 1 S(2)   &2.15421     & 2.17423    & $2786 \pm 4$  & $24 \pm 4$ & $0.23 \pm 0.01$ \\
H$_2$ 2 - 1 S(1 )  &  2.24772    & 2.26867   & $2794 \pm 4$  & $39 \pm 4$  & $0.51 \pm 0.02$ \\
\ion{He}{1}            & 2.05869    &  2.07807    & $2822 \pm 1$  & $33 \pm 1$ & $10.2 \pm 0.1$ \\
\ion{He}{2}\tablenotemark{f} & 2.03788   & 2.05654  & $2745 \pm 3$  & unresolved  &  $0.36 \pm 0.02$
\enddata
\tablenotetext{a} {measured with $1.1''$ diameter aperture}
\tablenotetext{b} {The listed uncertainties are from the rms noise and fitting the line profiles.}
\tablenotetext{c} {The uncertainty from the wavelength calibration is $\pm 12$ \kms.}
\tablenotetext{d} {velocity dispersion $\sigma_v$ corrected for instrumental line-width of 22 \kms\  at Br$\gamma$}
\tablenotetext{e} {The flux $F$ integrated over the line in units of $10^{-15}$ erg s$^{-1}$ cm$^{-2}$.
     It has not been corrected for extinction. Uncertainty from calibration and sky subtraction is about 10\%.}
\tablenotetext{f} {The centroid and $\sigma_v$ of the H$_2$ 1 - 0 S(2) line may have been affected by the residuals
from subtracting a neighboring night sky line.  The \ion{He}{2} line is partly contaminated by a night sky line.}
\end{deluxetable*}

\subsection{The Core in Near-Infrared Spectral Lines}

 The spatial resolution of the {\it K}-band observations is $\sim 0.5''$ or better,
 so somewhat better than that of the radio and mm-wave images.
The distributions of  Br$\gamma$, \ion{He}{1}, and {\it K}-band continuum
emission have a FWHM  of $1.1''$ (=  190 pc), which
 exceeds the  $0.6'' \times 0.4''$ size of the 0.6 Myr central star complex. Table\,\ref{IRTable}
 lists the detected {\it K}-band emission lines and their properties measured with a $1.1''$ diameter
aperture.  The H$_2$ rest wavelengths are from the wavenumbers in \citet{black87}.
A single Gaussian extrapolation to a diameter of $2''$
would increase the values of the flux $F$ by a factor of 1.18, compared to those  in this Table.
Although there is slightly more extended emission, this should be a reasonable approximation to the
total flux. The listed  values of the velocity dispersion $\sigma_v$  have been corrected for  instrumental line-width.
  The Br$\gamma$ and \ion{He}{1} lines have  $\sigma_v$ = $35 \pm 1$ \kms, and the \Hmolec\ line
has  $\sigma_v$ = $27 \pm 1$ \kms.
Thermal broadening  at an electron temperature
$T_e$ would produce a  $\sigma_v$  of only $12 \times (T_e/10^4  {\rm K})^{1/2}$  \kms.
Massive Oe stars and WN stars have line-widths that are an order of magnitude greater than the measured
Br$\gamma$ line-width of Feature i, but  winds from such stars could help power  turbulence
and expansion of the ionized gas in the core.  As a comparison, the SSC B1 in the Antennae overlap
region has  Br$\gamma$ and \Hmolec\ velocity dispersions of $ 45 \pm 4$ \kms\ and  $22 \pm 7$ \kms,
respectively
\citep{herrera17} for a region with a Br$\gamma$ FWHM size of 70 pc.  Thus the Br$\gamma$ velocity
dispersion of the Feature i core is somewhat less than for SSC B1 in the Antennae but both have similar values of
 the \Hmolec\ line-width.

\begin{figure}
\epsscale{1.0}
\plottwo{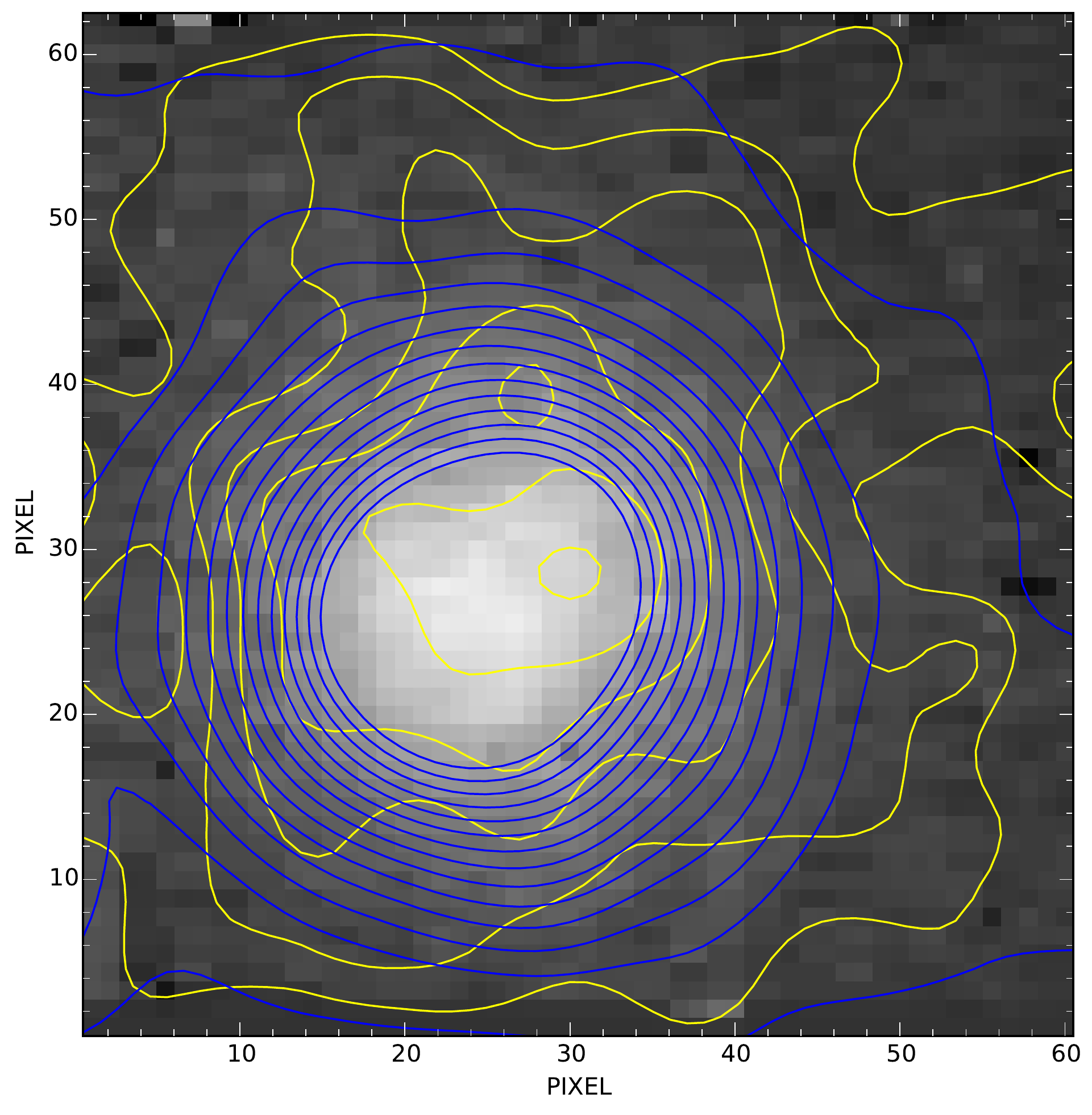}{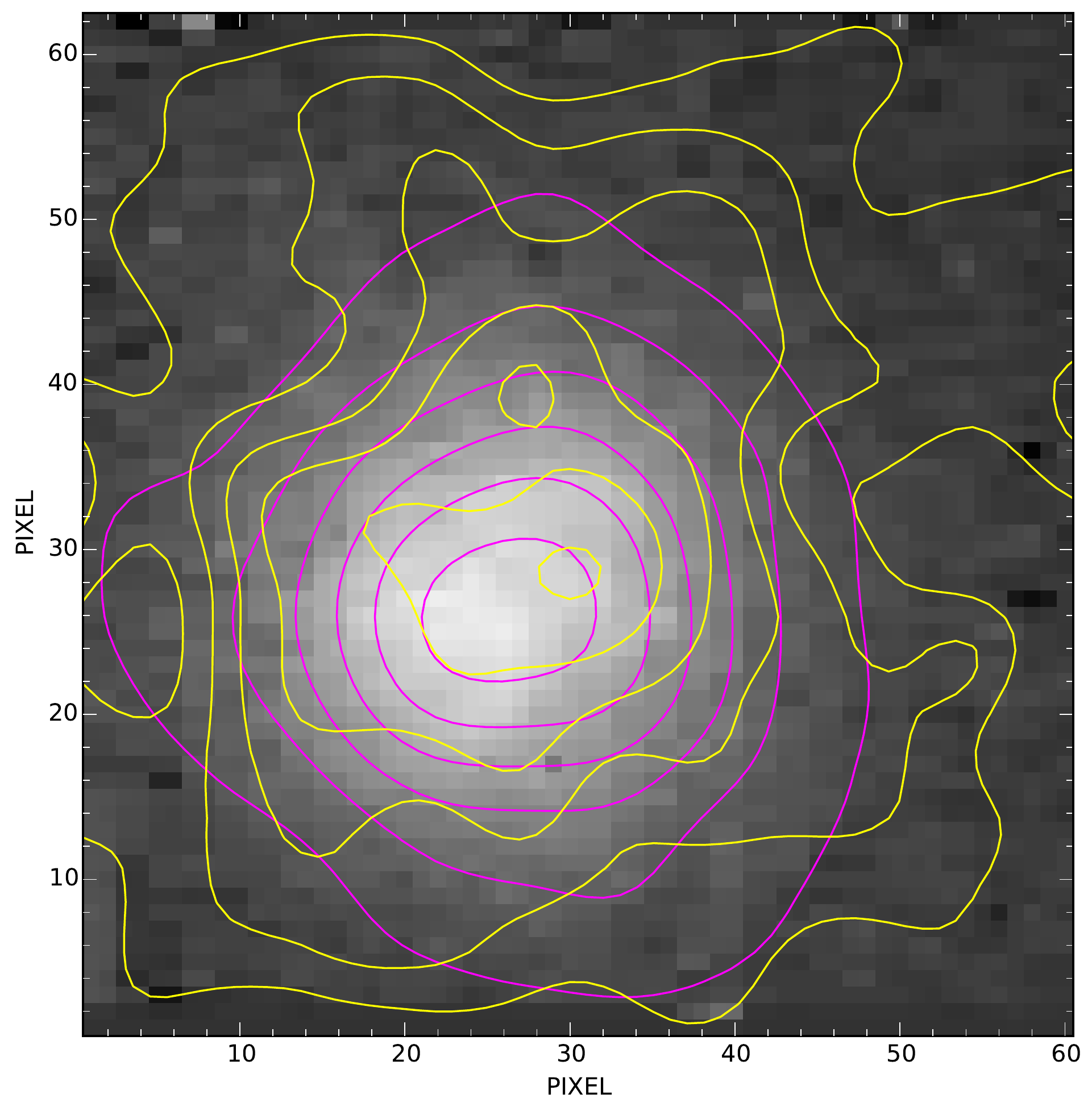}
\caption{Surface brightness images of near-IR emission from the Feature i core.
 Top: contours  of  Br$\gamma$ in blue and \Hmolec\ in  yellow overlaid on Br$\gamma$ in greyscale.
Bottom: contours of {\it K}-band continuum in magenta and \Hmolec\ in  yellow overlaid on
Br$\gamma$ in greyscale.
 The location of maximum surface brightness in \Hmolec\ is displaced a minor amount, $\sim 0.15''$,
northwest of the
 Br$\gamma$ peak, but aside from this there is no strong difference between the
spatial distributions of  \Hmolec, Br$\gamma$ emission, and {\it K}-band continuum.
In both panels,
the lowest contour is $3 \times$ the rms noise measured in the corners of the image.
The pixel size is $0.05'' \times 0.05''$, and the resolution is $\sim 0.5''$ or better.
\label{IRcontours}}
\end{figure}

Figure\,\ref{IRcontours} presents in greyscale the Br$\gamma$ image rebinned to  $0.05'' \times 0.05''$
pixels  with contours of Br$\gamma$, \Hmolec, and {\it K}-band continuum overlaid.
 The images have been smoothed with a 3 pixel smoothing kernel.
 From the  Gaussian fits
to the surface brightness distribution,  the
 location of maximum surface brightness in \Hmolec\ is displaced a minor amount, $\sim 0.15''$,  northwest of the
 Br$\gamma$ peak, but aside from this there is no strong difference between the
spatial distributions of  \Hmolec, Br$\gamma$ emission, and {\it K}-band continuum.

\begin{figure*}
\epsscale{0.95}
\plotone{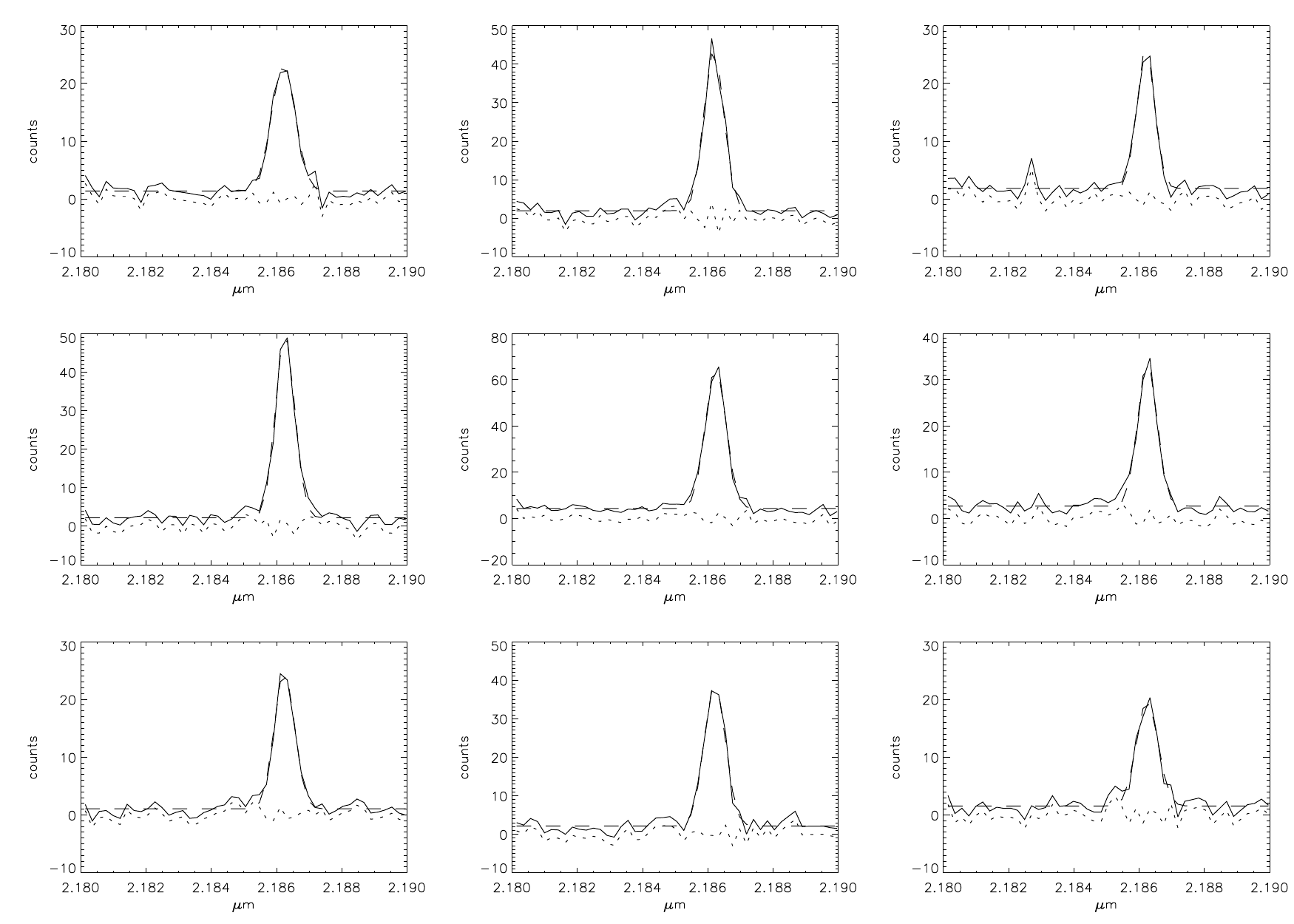}
\caption{  Br$\gamma$ line profiles and  residuals after subtracting a single
Gaussian plus a flat continuum fit at each position. The central panel is
located at the Br$\gamma$ peak, and the panels are spaced $0.5''$ apart. The solid curve is
the observed profile, the dashed cuve is the fit, and the dotted curve, the residuals.
 The residuals
 provide no evidence of a one-sided outflow.
\label{IRoutflow}}
\end{figure*}

We consider whether the ionized gas is bound to the central star complex.
 To estimate the escape velocity $v_{\rm esc}$, we adopt
from \citet{mengel02} the Virial relation

\begin{equation}
    v_{\rm esc} = [\frac{2 GM_c}{\eta r_{hp}}]^{1/2},
\end{equation}
where $r_{hp}$ is the projected half-light radius, the correction factor $\eta$ = 6 - 10 depends on the
mass distribution, and $M_c $ is the cluster mass.  For the central
star complex we take $r_{hp}$ as the geometric mean radius = 42 pc. The cluster mass $M_c$ is
the sum of  the stellar mass, $2 \times 10^6$ \msun, plus the mass of ionized gas within the star complex.
In Section 6.3  below, we assume a Gaussian distribution of electron density in the Feature i core and find
the total mass of ionized gas in the core  $\simeq 1.7 \times 10^7$ \msun. This is an upper limit to the
mass of ionized gas within the star complex, so $M_c < 1.9 \times 10^7$ \msun, and thus
$v_{\rm esc} <  62$ \kms/$\eta^{1/2}$ = 20 - 25 \kms, which is somewhat less than the observed
 Br$\gamma$ $\sigma_v$ of 35 \kms.
 We attribute the high velocity dispersion in the {\it K}-band spectral lines to a combination of high turbulence plus
some expansion of the ionized gas.
The shape of the Br$\gamma$ line profiles provides no evidence of a one-sided outflow.
 Figure\,\ref{IRoutflow} displays the Br$\gamma$ line profile, the Gaussian fit, and the
residuals after subtracting the Gaussian  fit at each position.
The central panel is
located at the Br$\gamma$ peak, and the panels are spaced $0.5''$ apart. This provides an
upper limit of 22 \kms\ (the instrumental line-width) to a nonsymmetric outflow but
does not rule out a spherical expansion of the ionized gas.
Gaussian fits to the Br$\gamma$ line profile at each position in the field show no systematic
change in centroid wavelength or in line-width across the region, except that the line-widths may
be slightly broader at the outskirts of the region where the S/N is worse.

We compare the measured value of the flux $F({\rm Br}\gamma$) =
$(2.0 \pm 0.2) \times 10^{-14}$ erg s$^{-1}$ cm$^{-2}$,
uncorrected for extinction,
with that expected from ionized gas associated with the 0.6 Myr central star complex.
Models by \citet{sternberg03} predict a Lyman continuum rate $N_{\rm Lyc}$ =
$2 \times 10^{53}$ s$^{-1}$ for an instantaneous burst of star formation with cluster  mass
$2 \times 10^6$ \msun\  at age 0.6 Myr, i. e., the values of mass and age of the  central complex
that \citet{elmegreen17} find by adopting the same IMF
(Salpeter with $M_{\rm L}$ = 1 \msun, $M_{\rm U}$ = 120 \msun) as  in \citet{sternberg03}.

For the 35 Mpc distance of NGC 2207,  the expressions in \citet{condon92} give

\begin{equation}
    N_{\rm Lyc} = 1.2 \times 10^{67} (T_e/10^4 K)^{0.31}  F({\rm Br}\gamma )      ,
\end {equation}
where  $F({\rm Br}\gamma$) in erg s$^{-1}$ cm$^{-2}$ is extinction corrected, and no Lyman continuum
photons are absorbed by dust or escape.  With the above measured value of  $F({\rm Br}\gamma$),
which is not corrected for extinction,
$N_{\rm Lyc}$ = $2.4 \times 10^{53}$ s$^{-1}$ for $T_e$ = $10^4$
or $1.9 \times 10^{53}$ s$^{-1}$ for $T_e$ = 5000 K.
Thus most of the observed Br$\gamma$ flux may result from Lyman continuum
photons emitted by the 0.6 Myr central star cluster. With $A_v$ equal
 3.3 mag for this star complex, then $A_K$ = 0.112 $A_v$ would be 0.37 mag (a factor of 1.4
in flux).
In the optical {\it HST} images, the central star complex is the only star complex in the
core, but there may be other ionization sources hidden behind the dark cone.

\begin{deluxetable*}{lcll }
\tablewidth{0pt}
\tablecaption{Gaussian Fits to the Core of Feature i
\label{Table6}}
\tablehead{
   \colhead{Data}   &  \colhead{$S_\nu$}              & \colhead{deconvolved size}    & \colhead{PSF}\\
                             &                                             & \colhead{(FWHM, PA)}   & \colhead{(FWHM, BPA)}\\
                            &    \colhead{(mJy)}                    &
}
\startdata
1.49 GHz\tablenotemark{a,b}   &3.4             & $\sim 1''$       & $1.92'' \times 1.11''$, $9\degr$       \\
 4.86 GHz\tablenotemark{a,c}  &1.4            &  $\sim 1''$       & $1.93'' \times 1.04''$, $169\degr$     \\
4.86 GHz\tablenotemark{d}   & $2.66 \pm 0.02$  & $1.20'' \pm 0.03'' \times 0.82'' \pm 0.03''$, $43\degr  \pm 4\degr$
   & $2.47'' \times 1.30''$, $8\degr$  \\
88 GHz    & $1.50 \pm 0.16$ & $0.9'' \pm 0.2'' \times 0.8'' \pm 0.2''$, $155\degr \pm 44\degr$
  & $1.26'' \times 1.00''$, $-81.7\degr$  \\
100 GHz  & $1.45 \pm 0.07$   & $1.03'' \pm 0.05'' \times 0.73'' \pm 0.06''$, $8\degr \pm 10\degr$
  & $1.07'' \times 0.85''$, $-83.5\degr$   \\
106.4 GHz   & $1.52 \pm 0.12$  & $1.2'' \pm 0.2'' \times 0.6'' \pm 0.3''$, $167\degr \pm 16\degr$
  & $1.64'' \times 1.22''$, $-79.7\degr$   \\
8 \micron\   & $14.3 \pm 0.02$  & $1.07'' \pm 0.03'' \times 0.70'' \pm 0.04''$, $17.5\degr \pm 0.4\degr$
  & $2.4''$
\enddata
\tablenotetext{a} {VLA scaled-array pair observations in 1986 by \citet{vila90} }
 \tablenotetext{b} {VLA radio continuum observations on 18 March 1986}
\tablenotetext{c} {VLA radio continuum observations on 16 August 1986}
\tablenotetext{d} {VLA radio continuum observations on 14 April 2001}
\end{deluxetable*}

Generic models by \citet{black87} of the excitation of H$_2$  lines
[see also discussion by \citet{rosenberg13} and \citet{doyon94}]
predict the  flux ratios 2-1 S(1)/1-0 S(1)  and 1-0 S(0)/1-0 S(1) should have values  $\sim (0.53 - 0.56)$
and $\sim  0.46$, respectively,  for excitation by UV radiation and
$\sim 0.1$ and $\sim 0.2$ for excitation by collisions in shocks.
For the H$_2$ lines  in the core of Feature i, the observed flux ratio  of 2-1 S(1)/1-0 S(1) equals
$0.34 \pm 0.02$ and
the observed flux ratio of 1-0 S(0)/1-0 S(1) equals  $0.52 \pm 0.02$. These values
  indicate  that excitation by  UV radiation  rather than
by collisions in shocks is probably dominant in the core.
The  similarity between the
spatial distributions of \Hmolec\ and Br$\gamma$ emission supports this interpretation.

The flux ratio of  \ion{He}{1}  2.06 \micron\ to Br$\gamma$ is an indicator of the effective
temperature $T_{\rm eff}$ of the stars producing the ionization, although the ratio is also
sensitive to the electron density $n_e$,  the dust content, and the geometry of the \HII\
region \citep{doherty95}.  In the Feature i core, this ratio has a value of  0.52, which is the
maximum value  in the model by \citet{shields93} for $n_e$ = $10^2$ cm$^{-3}$,  and for that value of
$n_e$ implies $T_{\rm eff}$ = $4 \times 10^4$ K \citep{doherty95}.
Although partly contaminated by a night sky line, the presence of  the
  \ion{He}{2}\   2.04 \micron\  line in  the Feature i spectrum is indicative of  high $T_{\rm eff}$.

\subsection{Radio and mm-wave Continuum}

Table\,\ref{Table6}  lists the results of  two-dimensional Gaussian fits to the continuum emission from
the Feature i core at  various frequencies.
 For the 4.86 GHz radio continuum observations in 2001
and the {\it Spitzer} 8  \micron\ observations, both of which exhibit a bright core embedded
extended emission, the Gaussian fits employ the sum of two Gaussians plus a flat baseline, and only
the Gaussian characterizing the  core is listed here. For the rest of the data sets, the fit consists
of a single Gaussian plus a flat baseline.  The core of Feature i is the  brightest source in
NGC 2207/IC 2163 in the 100 GHz continuum, radio continuum, and 8 \micron\ emission,
and all of the measurements in Table\,\ref{Table6} get about the
same value for its deconvolved FWHM, $1.0'' \times 0.7''$  (170 pc $\times$ 120 pc).
In the 100 GHz continuum, the core has a
 major axis PA of  $8\degr \pm 10\degr$, which
  differs considerably from the $- 45\degr$ major axis PA
of the 0.6 Myr central star complex in the {\it HST} images.

After convolving the 88 GHz, 100 GHz, and 106 GHz continuum images to the same
synthesized beam, we compare their  flux densities
for the Feature i core
and  find  an 88 GHz to 106 GHz spectral index
   $\alpha = - 0.2 \pm 0.7$.  This rules out
optically-thin heated dust which would have $\alpha = 3.8$ \citep{scoville14} and
free-free emission that is optically thick at 100 GHz (which would have $\alpha = 2$)
as major source(s) of
the emission at 100 GHz.  We infer that at 100 GHz the main contributor to  the flux
density of the core  is free-free emission that is optically-thin at 100 GHz plus a
contribution from nonthermal emission.

 We need to reconcile the radio continuum and 100 GHz continuum flux densities
of  the Feature i core  with the Br$\gamma$ flux.
Based on the relations in \citet{condon92} between optically-thin free-free flux density $S_{\rm ff}$
and $F({\rm Br}\gamma$) for \HII\ regions,  we find that if $T_e$ = 5000 K
the measured value of  $F({\rm Br}\gamma$) uncorrected for extinction corresponds to
$S_{\rm ff}$(100 GHz)  = $0.95 \pm 0.10$ mJy.  Allowing for some extinction at Br$\gamma$,
this is the minimum to be consistent with $F({\rm Br}\gamma$), and thus
at least 67\%  of the 100 GHz flux density from the Feature i core needs to be
optically-thin free-free emission.
We rule out  $T_e \geq 10^4$ K as otherwise the total $S$(100 GHz)
of the core would be too small to account for the measured Br$\gamma$ flux.

 In Table\,\ref{Table6}  the 1986 radio continuum observations by \citet{vila90}
are a VLA scaled-array pair.  Since for the core they measured  $S_\nu$(4.86 GHz) = 1.4 mJy
and a radio spectral index $\alpha$ = $-0.7$ (whereas optically-thin free-free emission would
have $\alpha$ = $-0.1$), the issue is how to get $S_{\rm ff}$(100 GHz) $\sim 1$ mJy.
We propose two types of ad hoc solutions: either (a) short term variability of the
nonthermal radio component with a decrease in the nonthermal component
between March 1986 (when the 1.49 GHz emission was
observed) and August 1986 (when the 4.86 GHz emission was observed),
or (b) the presence of a component
whose free-free emission is optically thick (and thus largely hidden)
at 4.86 GHz but optically thin at 100 GHz.  The actual explanation may be a combination
of (a) and (b).   In both cases, the nonthermal emission at 100 GHz may
contain a contribution from
 nonthermal sources  not detected at 4.86 GHz because of  absorption.

Evidence for radio continuum variability of the core  is the  1.26 mJy increase  in its
$S_\nu$(4.86 GHz) between 1986 and 2001. Although
this may have been a radio SN  \citep{kaufman12}, it could have been some
other type of  energetic outburst.
  In Section 4.3, we argued that the possible
  ULX detected by \citet{mineo14}  nearly coincident with  the radio continuum peak
could be an
intermediate black hole because of the high extinction at that location. If so, it could serve
as a source of short-term radio variability.

 If variability of the  nonthermal radio component  in 1986 is the answer, we present the
following
example of how to satisfy $S_{\rm ff}$(100 GHz) $\sim 1$ mJy.
We assume that the free-free emission is optically-thin for $\nu \geqq 1.49$ GHz.  If
 $S_{\rm ff}$(4.86 GHz) were 1.24 mJy, then
 $S_{\rm ff}$(1.49 GHz) would be 1.40 mJy, and
$S_{\rm ff}$(100 GHz) would be 0.92 mJy,  close
to the minimum free-free flux density necessary if $T_e \sim$ 5000 K.
Then with an adopted spectral index $\alpha_{\rm nth}$ = $-0.9$ for the nonthermal component,
this would require
the value of $S_\nu$(1.49 GHz) to have decreased from the observed 3.4 mJy in
March 1986 to 1.9 mJy in August 1986 and the value of $S_\nu$(4.86 GHz) to have
decreased from  2.1 mJy in March 1986 to the observed 1.4 mJy in August 1986.
Approximately the same changes in $S_\nu$(1.49 GHz) and $S_\nu$(4.86 GHz) between
March and August would suffice to give $S_{\rm ff}$(100 GHz) = 0.92 mJy
 if  $\alpha_{\rm nth}$ were $-0.7$, or we could allow
$\alpha_{\rm nth}$ to vary between the two dates.

 Alternatively, if the nonthermal radio flux densities did not vary between
March 1986 and August 1986,
we propose that some part of $S_{\rm ff}$(100 GHz)  arises from a component
that is optically thick at 4.86 GHz but optically thin at 100 GHz.
We  extrapolate the flux densities from 4.86 GHz to 100 GHz by assuming
 the measured $S_\nu$(100 GHz) of the core has the following components:

\begin{equation}
    S_\nu = S_{\rm nth} + S_{\rm ff,diffuse} +S_{\rm ff,dense},
\end {equation}
where  $S_{\rm nth}$  is nonthermal emission,
  $S_{\rm ff,diffuse}$  is emission
that is optically-thin free-free at 4.86 GHz and at 100 GHz, and  $S_{\rm ff,dense}$ is
from an hypothesized  group of  ultra-dense \HII\ regions (UD HII regions) that are optically-thick
at 4.86 GHz but optically-thin at 100 GHz, with
turnover frequency $\nu_{\rm t,dense}$ (at which $\tau_{\rm ff,dense}$ =1).   \citet{kobulnicky99} use
the term UD HII region for the massive analog of an ultra-compact \HII\ region.
The UD HII regions in Feature i would be analogous to  the four UD HII regions in Henize 2-10
studied by \citet{johnson03}.

Table\,\ref{example}  provides  examples yielding $S_{\rm ff} $(100 GHz) = 1 mJy for two
different choices of $S_{\rm ff,diffuse}$(4.86 GHz),  0.7 mJy and 0.3 mJy
[50\% and 21\%, respectively, of the observed $S_\nu$(4.86 GHz) in 1986] and  two different
 choices of  $\nu_{\rm t,dense}$, 10 GHz and 20 GHz.
 For free-free emission, the  optical depth $\tau_{\rm ff} \propto  \nu^{-2.1}a(\nu,T_e)$
[where the factor $a(\nu,T_e)$ is  $\sim 1$ from \citet{mezger67}]
and $S_\nu \propto \nu^2 [1 - \exp(-\tau_{\rm ff})]$.
This gives the values of
 $\tau_{\rm ff, dense}$(4.86 GHz), $S_{\rm ff,dense}$(4.86 GHz), and
the emission measure of the dense component $EM_{\rm dense}$ listed in Table\,\ref{example}.
  Subtracting the diffuse and dense components of
the free-free emission from the total flux densities measured in 1986 produces the  values of
$S_{\rm nth}$(4.86 GHz) and $\alpha_{\rm nth}$.
 For the set of examples in Table\,\ref{example}, the nonthermal emission
detected at 4.86 GHz in
1986 contributes, at most, 4.6\% of the measured $S_\nu$(100 GHz).

\begin{deluxetable}{lllll}
\tablewidth{0pt}
\tablecaption{Examples giving $S_{\rm ff}$(100 GHz) = 1 mJy with $T_e$ = 5000 K
\label{example}}
\tablehead{\colhead{Property} & & \colhead{Values}
}
\startdata
$S_{\rm ff,diffuse}$(4.86 GHz) (mJy)    & 0.70    & 0.70  & 0.30  & 0.30 \\
$\nu_{\rm t,dense}$  (GHz)                  & 10      & 20     &  10    & 20 \\
$S_{\rm ff,diffuse}$(100 GHz)  (mJy)    & 0.52    & 0.52   &  0.22  &  0.22 \\
$S_{\rm ff,dense}$(100 GHz)  (mJy)     & 0.48   & 0.48   & 0.78   &  0.78 \\
$\tau_{\rm ff, dense}$(4.86 GHz)  & 4.64        & 20.5        & 4.64        & 20.5 \\
$S_{\rm ff,dense}$(4.86 GHz) (mJy)      & 0.16  & 0.04   & 0.26  & 0.06 \\
$EM_{\rm dense}$  ($10^8$ pc cm$^{-6}$)  & $1.6$  & $6.9$
                &  $1.6$  & $6.9$ \\
$S_{\rm nth}$(4.86 GHz)\tablenotemark{a}  (mJy)  & 0.54 & 0.66  & 0.84  & 1.04\\
$\alpha_{\rm nth}$                        &  $-1.3$   & $-1.2$    &  $-1.1$    & $-0.91$
\enddata
\tablenotetext{a}{in 1986}
\end{deluxetable}

If the $S_{\rm ff,dense}$(4.86 GHz) were from
 a single spherical UD HII region with uniform electron density $n_e$, then
for  the Table\,\ref{example}  examples, the UD HII region would have a
radius of  5.3 pc - 2.0 pc and a
proton column density $N_p$ of $(8 - 13) \times 10^{22}$ cm$^{-2}$.
Unless the dust-to-gas ratio is low within the UD HII region,  the large values of
 $N_p$  imply the UD HII
 would have appreciable extinction  in {\it K}-band.
The free-free optical depth  depends differently on density than the visual
extinction does.  Hence  a group of  UD HII regions that
are flattened along line-of-sight with a nonuniform electron density and with the same values of  $EM_{\rm dense}$
and $\tau_{\rm ff, dense}$(4.86 GHz)  as in Table\,\ref{example}
would have a lower {\it  K}-band extinction than  the single uniform sphere example.

To calculate the total mass of ionized gas in the core, we take $
S_{\rm ff}$(100 GHz) $\sim 1$ mJy for the core,
  assume  a Gaussian distribution of electron density, and  apply Equation A.14
  of \citet{mezger67}. The result is
 an \HII\  mass of  $\sim 1.7 \times 10^7$ \msun\ for the core.

We offer the following comment about warm dust emission at 100 GHz.
\citet{yang07} obtain a median dust temperature $T_d$ = $39 \pm 8$ K
for 18 nearby LIRGs, and \citet{wild92} get $T_d$ = 45 K for the bulk of  the ISM in the
M82 starburst.
Suppose  warm dust  contributes  7\% (i.e., 0.1 mJy) of the measured $S$(100 GHz)
of the Feature i core with $T_d$ = 45 K  for the bulk of the ISM.  Scaling
Equation (12)  of \citet{scoville14}
to $T_d$ =45 K then gives  a  core ISM mass $M_{\rm ISM}$ of $7 \times 10^7$ \msun.
 For the core, the CO flux density implies a molecular mass
  $M({\rm H_2})$ after including helium of
 $3.4 \times 10^7$ \msun\ with a $2.4''$-diameter aperture (twice the
 geometric mean of the CO HPBW), and part of this molecular mass resides in the dark cone,
so not in the disk. Therefore an upper limit to the sum of the molecular plus
\HII\ mass of the core is $\sim 5 \times 10^7$ \msun.  Either  the contribution of warm dust emission
to $S$(100 GHz) is less than 0.1 mJy  or
the \HI\ mass  of the core is at least comparable to its \HII\ mass.

\begin{figure}
\epsscale{0.82}
\plotone{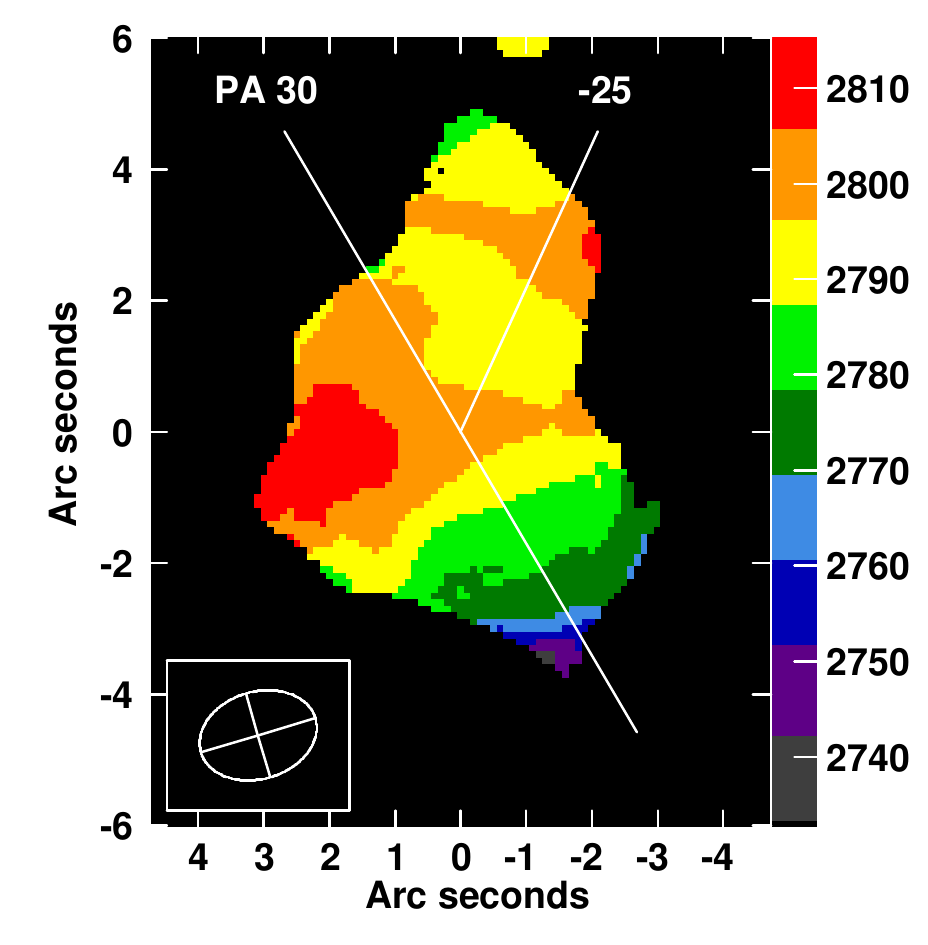}
\caption{CO velocity field of Feature i  from the CO Combined data.
  The  display  is centered on the radio peak, and the
 axis of the dark dust cone is represented by the line drawn at
 PA = $30\degr$. The extended  8 \micron\ emission from Feature i
has its major axis at PA = $-25\degr$, which is also the PA of the CO lobe north-northwest
of the core.
\label{fig6}}
\end{figure}

\begin{figure}
\epsscale{1.0}
\plotone{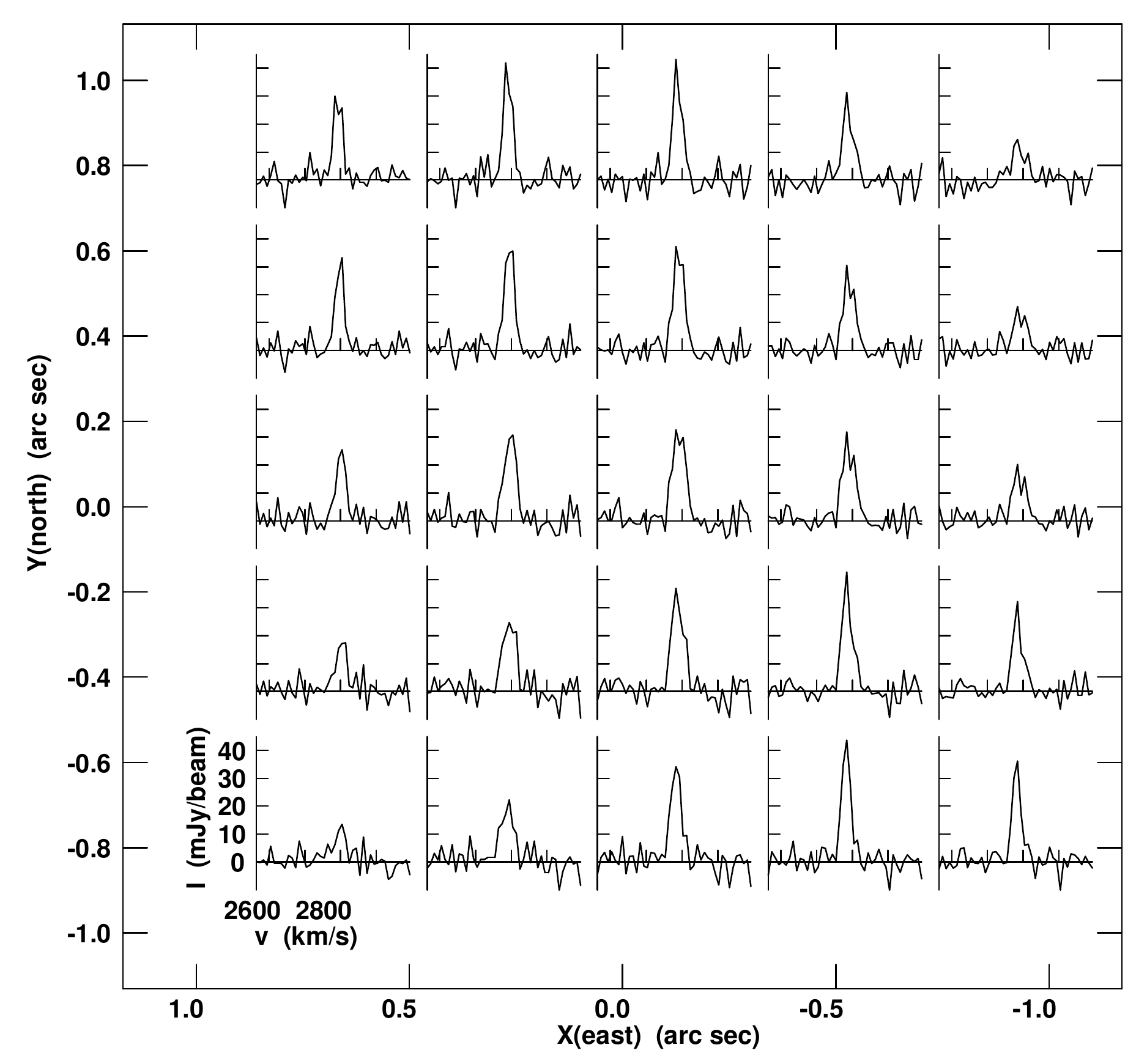}
\caption{CO spectra of the core, spaced $0.4''$ apart, from the CO Cycle 2 data.
The central panel  is at  the radio continuum peak. The noise $\sigma_{\rm eff}$ = 3.8 \mJybeam, equivalent to 0.25 K.
Some of the line profiles are asymmetric, indicative of more than one component or outflow.
\label{fig7}}
\end{figure}

\section{The Dark Dust Cone and Internal Kinematics of Feature I in Cold  Gas}

\subsection{Measured Velocities}

We  consider whether there is kinematical evidence for the suggestion in
Section 5 that the CO emission along the opaque dust cone
is from gas outflowing perpendicular to the disk at Feature i.
To get this gas in front of the midplane (relative to us), it needs
to have a velocity component  in our direction, and thus  the CO emission from the
dark cone should be blueshifted relative to  the rest of Feature i.

 Figure\,\ref{fig6} displays  the intensity-weighted  velocity field (first moment image)
from the CO Combined data.  The channel width is 4.87 \kms; this implies a
velocity uncertainty  of $\pm 2.4 $ \kms.
 Because Feature i is a two-lobed source in CO, the velocity field
is complex.   Figure\,\ref{fig6} reveals that
the velocity along the dark cone
is blueshifted relative to the rest of Feature i.
 Using a $1.8''$-diameter aperture
(to approximate the synthesized beam area) centered along the dark cone (PA = $210\degr$)
at a distance of  $2.0''$
from the radio peak,  we find the mean value
of $v$ is blueshifted by 18 \kms\  relative to its value at the radio peak.

\begin{figure}
\epsscale{1.0}
\plottwo{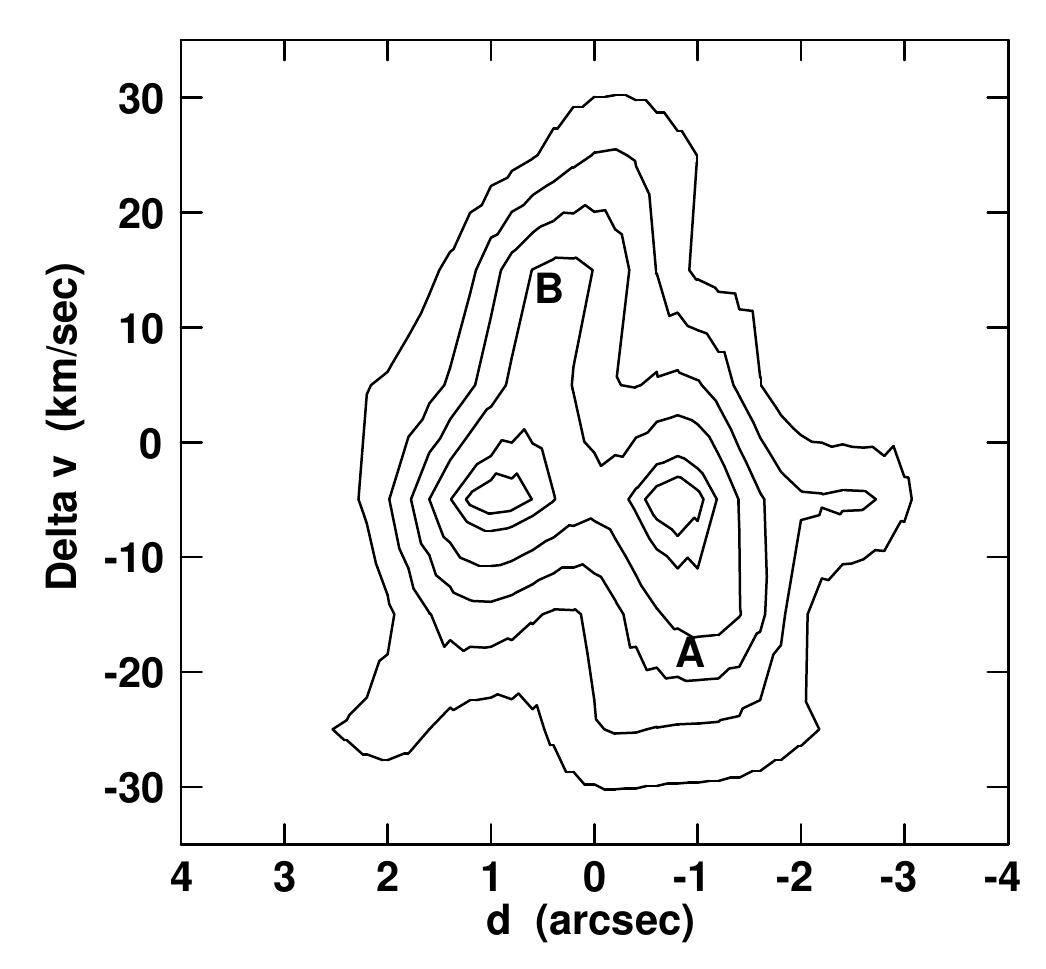}{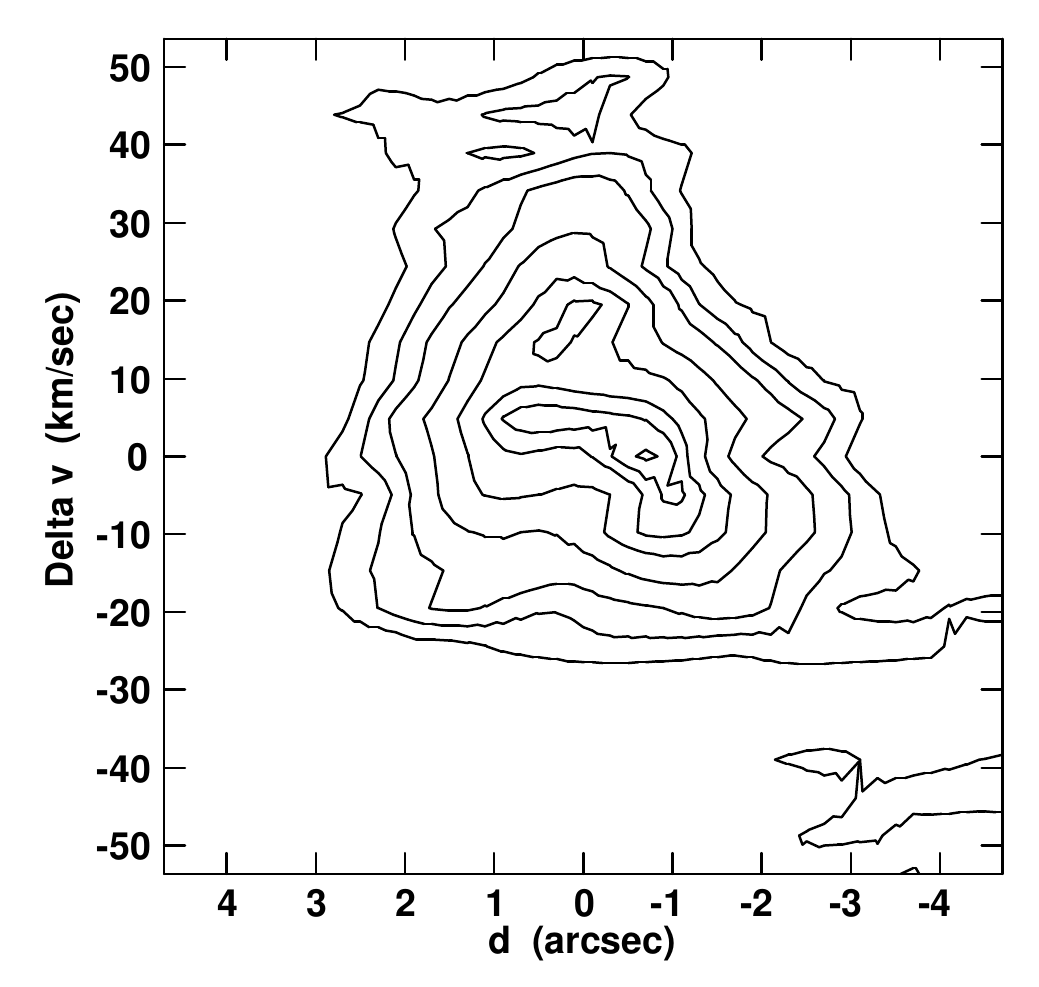}
\caption{ Position-velocity diagrams.  In both panels, the center is at the location of
the radio peak,  $\Delta v$  =0 corresponds to a velocity  of 2789 \kms,
 and the abscissa is along PA = $30\degr$. Negative values of  the
displacement $d$ are on the dark cone; positive values of $d$ are on the
opposite (northeast) side of the radio peak.
Top: From the CO Cycle 2 cube with contours at
2, 4, 6, 8, 10, 11 $\times \sigma_{\rm eff}$,
where $\sigma_{\rm eff}$ = 3.8 \mJybeam\ = 0.25 K.
 Label A identifies an approaching outflow component on the dark cone, and
Label B, a receding outflow component on the northeastern side of the radio peak.
 Bottom: From the CO Combined cube with contours at
 2, 4, 6, 9, 12, 14, 15, 16 $\times  \sigma _{\rm eff}$,
where $\sigma_{\rm eff}$  = 3.3 \mJybeam\ = 0.13 K.  More sensitive to extended
emission, it reveals that CO emission from the dark cone is present to $d$ = $- 3''$.
\label{fig8}}
\end{figure}

Figure\,\ref{fig7} displays \CO\ spectra of the core from the unmasked
 CO Cycle 2 data.
The central panel is located at the radio continuum peak, the panels are spaced $0.4''$ apart,
and each individual line-profile is an average over a $0.4'' \times 0.4''$ box.
Some of these line profiles,
exhibit  an asymmetry or   hint of more  than one component.
Thus some of  the velocity dispersion in the cold molecular gas may result from outflows or a blending of
components rather than turbulence.
From a Gaussian fit to the CO line profile at the radio peak,
 the FWHM line-width corrected for the velocity resolution is
 $42 \pm 1$  \kms\  [i.e., $(\sigma_v)_{\it corr} = 18 \pm 1$ \kms].
[With the  unmasked CO Combined data, the Gaussian fit gives a slightly
greater value of $47 \pm 1$ \kms\ for the corrected FWHM line-width.]
In the velocity dispersion  (second moment) image, the velocity dispersion $(\sigma_v)_{\it corr}$
at the radio peak is 15 \kms\  and in the $2'' \times 2''$ box displayed in this
figure, the average $(\sigma_v)_{\it corr}$ =$14 \pm 1$ \kms.  Recall from Section 6.2 that
Br$\gamma$ and the warm molecular gas traced by the infrared \Hmolec\  lines have
greater line-widths  than this with $\sigma_v$ = 35 \kms\ and 27 \kms,
 respectively, and  that the spatial distribution of ionized and warm molecular gas
is quite different from that of the cold molecular gas.

Figure\,\ref{fig8}  displays  position-velocity diagrams: the top panel is  from the
 CO Cycle 2 data and the bottom panel from the CO Combined data. Both are
 centered on the radio peak and oriented along  the dark cone
at  PA = $30\degr$.
 On the abscissa, negative values of the plane-of-sky
displacement $d$ are on the dark cone; positive values of $d$ are on the opposite side of
 the radio peak.
 The CO Combined cube in the bottom panel
 reveals that the CO emission along the
dark cone is present to  $d$ = $- 3''$, consistent with the length of the visually opaque cone.

 Because of  its higher spatial resolution, the top panel of   Figure\,\ref{fig8}  provides
important detailed information, and the rest of our discussion here is based on it.
 Its two brightest emission knots
are $I$(CO) = 46 \mJybeam\ at $d$ = $- 0.8''$ and  $I$(CO) = 45 \mJybeam\ at $d$ = $1.0''$.
Each of these two molecular clumps bounding the ionized region in the core
has a velocity of  2784 \kms, which we adopt as the disk velocity.

Most of the emission along the dark cone (negative values of $d$)
 is blueshifted relative to this disk velocity. There are two outflow components
to the CO emission from Feature i: (1) an approaching outflow component (labelled A
in Figure\,\ref{fig8})  along the dark cone, and (2) a  receding  outflow component
(labelled B)  on the opposite side of the radio peak. Relative to the disk velocity, Component A
is blueshifted by  11  \kms\  at the 8 $\sigma_{\rm eff}$
contour level and by 16 \kms\ at the 6 $\sigma_{\rm eff}$ contour level. For $d $ = 0 to $ - 2''$,
its ouflow velocity appears roughly constant. Component B
is redshifted by 20 \kms\  at the 8 $\sigma_{\rm eff}$  level and by 25 \kms\ at the
6 $\sigma_{\rm eff}$ level relative to the disk velocity.
Its outflow velocity decreases with distance from the center for $d$ =0 to $1.2''$.
Since, compared to the Component A,   Component B
does not extend as far and has a greater magnitude outflow velocity,   we infer it is from
 more recent outburst than A.
The difference in outflow velocities on the two sides of the radio peak explains why
the intensity-weighted velocity at the radio peak is somewhat greater than
the disk velocity based on  the position-velocity diagram.

There is other evidence of outflowing cold molecular gas on the far side of the Feature i core.
The detected HCO$^+$ emission lies roughly along the left fork
of the red V, i. e.,  on the opposite side of the radio peak
from the dark dust cone.
Restricting to positions where $I({\rm HCO}^+$) exceeds $(2 \sigma_{\rm eff}) \times
$(2 channel widths), we find that along the left fork of
the red V from $d$ = $0.3''$ to $1.3''$, the HCO$^+$ velocity field
 has
a redshift of  8 to 16 \kms\
relative to the CO disk velocity of 2784 \kms,
and thus HCO$^+$ is also participating
in a redshifted  outflow on the northeastern side of the radio peak.

 The southwestern lobe of CO emission (CO lobe II in Figure\,\ref{arm}) coincides with
and has about the same length as the visually opaque dust cone.
Attributing the blueshifted CO gas  (Component A) along the dark cone to an outflow
on the near side of the core  is the only way to get
enough cold molecular gas in front of the disk (relative to us) to explain
the high $A_{\rm v}$ of the opaque cone. For the redshifted gas
(Component B)  an outflow on the far side of the core is not the only
possible interpretation as there could be noncircular motions in the disk (e.g., due to the
spiral arm), but its
location diametrically opposite the dark cone (see Figure\,\ref{colorcontours})
suggests that it is also an outflow.

Our \HI\ data  have too low a spatial resolution ($13.5'' \times 12''$)
for measuring outflows from the Feature i core. Instead we compare \HI\ velocities
\citep{elmegreen95a}  and CO velocities (from the 2014 observations)
averaged over a $14''$-diameter aperture centered on the Feature i core:
 the mean  velocity $v \pm$ rms scatter   is $2769 \pm 11$ \kms\ in \HI\ and
$2786 \pm 15$ \kms\ in CO.
Feature i is located in the large area of NGC 2207 that has very broad and asymmetric
\HI\ line profiles.
Averaged over the $14''$-diameter aperture centered on Feature i,
 $(\sigma_v)_{\it corr}$,  the velocity dispersion
corrected for the velocity gradient across the HPBW,
is $52 \pm 4$ \kms\ in \HI\ and $13 \pm 8$ \kms\ in CO.

\subsection{Source of Outflows}

We consider what powered the outflow of cold molecular gas to create the optically-opaque
dust cone extending $3''$ from the center of the core.
On the assumption that supernovae and stellar winds from O and WR stars
integrated over their lifetimes
supply the necessary mechanical energy,
we estimate the SFR in the core required to produce the observed  line-of-sight
outflow velocities of the swept-up mass in the dark cone.
The 0.6 Myr central star complex produces ionization and contributes to expansion of the \HII\ region but
 is too young to provide most of  the mechanical energy of the outflow or to account for the observed
variable radio emission or  the length of the dark cone. Since it is too young to have hosted a supernova,
it is not  a source of the cosmic ray electrons necessary for the observed nonthermal radio emission.
The 0.6 Myr
star complex is the only star complex in the core detected in {\it B}-band,  but older star complexes (up to
60 Myr) are found elsewhere in Feature i.
 We conjecture that in the core there were previous
generations of star formation whose optical emission is now hidden behind the high extinction near
the radio continuum peak.

 We assume  the dark cone is perpendicular to the  Feature i  disk
and adopt the following  notation:
$i_{\rm feati} $ is the inclination  of the disk at Feature i,
 $H_z$ = $d /\sin i_{\rm feati}$ is the
perpendicular height of the dark cone above this disk,  $v_{\rm outflow}$ is
the observed line-of-sight component of the outflow velocity, and
 $v_z$ = $v_{\rm outflow}/\cos i_{\rm feati}$ is the component of the
outflow velocity perpendicular to this disk.
For the calculations below of the required mechanical energy, we adopt as representative values
$v_{\rm outflow}$ = -13 \kms\ on the dark cone and  +23 \kms\ on the opposite side of the radio peak.

\subsection{Required Mechanical Energy and SFR}

Simulations by \citet{ceverino12} of clumps with masses $10^8 - 10^9$ \msun\  in disk galaxies find
that because  angular momentum is approximately conserved during the clump's contraction,
 the spin vector of such a massive clump is generally aligned with the angular momentum vector of the
disk in the clump's neighborhood.
 Since the total mass in gas of Feature i lies in this range, their results should apply to Feature i.
If we neglect a possible warp of the NGC 2207 disk at Feature i and take $i_{\rm feati}$  =
$i$ of central disk of NGC 2207 = $35\degr$, then
$v_z$ = $- 16$ \kms\
on the dark cone and 28 \kms\ on the opposite side.
These are low velocity outflows, but the outflow velocities would have decreased as the
molecular gas traveled away from the sources because of  (a)  the local potential well of the
Feature i core, and (b) mass loading as the outflowing gas
entrains surrounding gas; $E_{\rm kin}$ of bulk motion $\sim M^{-1}$, where $M$ is the
swept-up mass.

To estimate the mass of the outflowing molecular gas in the dark cone, we restrict our measurement to
 emission with velocities $\leq$ 2774 \kms\ in the  CO Cycle 2 cube
and sum the emission in the quadrant southwest of the radio peak. After correcting for helium,
this gives  $M({\rm H}_2)$ = $8 \times 10^6$ \msun.  Of this mass, 36\%  lies  on the dark cone
at   $|d| \geq 1.4''$ from the radio peak,
25\% at $|d| \geq 2.0''$, and only 4\% at
$|d| \geq 2.8''$. With $M({\rm H}_2)$ = $8 \times 10^6$ \msun\ and $v_z$ =
16 \kms, the  estimated kinetic energy of bulk motion $E_{\rm kin}$ = $2 \times 10^{52}$ ergs for the dark cone.
For the redshifted outflow on the far side of the core, the value of $v_z^2$ is higher by
a factor 3, but it is less extensive and its mass is unknown. On the assumption
  that $E_{\rm kin}$ of the redshifted outflowing gas on the far side is comparable to that of
the dark cone, the
cumulative effect of supernovae and stellar winds in the core needs to produce a total  $E_{\rm kin}$  of roughly
$4 \times 10^{52}$ ergs to power both outflows.
Since  according to \citet{castor75} the amount of mechanical energy
injected into the ISM  by stellar winds  from a single high mass star integrated over its lifetime
is comparable to that from a core-collapse supernova, we attribute half of  this $E_{\rm kin}$ to
supernovae.

We consider whether the SFR in the core of Feature i can
achieve the required value of $E_{\rm kin}$.  Only about
0.1\% of the energy of a supernova remains as $E_{\rm kin}$ of bulk motions by the
time the material has reached a few tens of pc from the explosion \citep{martizzi15};
see also \citet{iffrig17}.
Therefore to get a $E_{\rm kin}$ of roughly $ 2 \times 10^{52}$
ergs in this kind
of wind from SNe requires at least $2 \times 10^4$
supernovae.  Since  it is hard to gather energy over more than
5 - 10 Myr because of the lifetimes of the progenitor stars of core-collapse supernovae and the crossing times,
we need to have at least $2 \times 10^4$ SNe
occur in a time interval less than 10 Myr, if both the blueshifted and the redshifted outflows
originated within the same 10 Myr period.
A standard  IMF gives about one supernova for every 100
\msun\ of stars formed, so we would  need a SFR $> 0.2$  \msun\ yr$^{-1}$ over 10 Myr to
account for both outflows   (or  0.1  \msun\ yr$^{-1}$ to account for just the dark cone).
Using a $10''$ diameter aperture, \citet{smith14} find the SFR of Feature i  is 1.7 \msun\ yr$^{-1}$
from the H$\alpha$ and 24 \micron\ luminosities (sensitive to the SFR over the past
10 Myr)  and 1.5 \msun\ yr$^{-1}$ from 24 \micron\ and $\it GALEX$ NUV (sensitive to the SFR over a
longer timescale).
The  distributions of mm-wave continuum, radio continuum, and 8 \micron\ emission
imply that a significant fraction
of this SFR occurs in the core,
 e. g., the core contributes 40\%  of the 8 \micron\  flux.
Thus it should be possible to satisfy the energy constraints
with star formation at the current rate  over  10 Myr.

We have not ruled out the possibility that nonthermal sources other than supernovae may also contribute  to the mechanical energy.

 For the CO outflow associated with the dark cone, CO emission is detected to a
   perpendicular height $H_z \sim 0.9$ kpc above the disk, and
the  molecular mass $M({\rm H}_2)$  at  $H_z > 600$ pc is $\sim 2 \times 10^6 $ \msun.
For this outflow  $H_z$ at $d$ = $- 2''$  is 590 pc,
 and the  time required for the molecular gas
to travel from $d$ = $-1''$ to $d$ = $-2''$ with a constant speed $v_z$ of 16 \kms\
would be 18 Myr.

  In the encounter model by \citet{elmegreen95b} which
 reproduces the observed S-shape of the isovelocity contours in NGC 2207,  the tidal
force exerted by IC 2163 causes a warp  of the outer disk of NGC 2207 with
  the northwestern part of NGC 2207  bent more towards us.
If at Feature i the disk is warped  with  $i_{\rm feati} > i$
of the NGC 2207 central disk,  then $v_z$ and the required $E_{\rm kin}$ would be greater
than in the unwarped disk case.
If, averaged over 10 Myr, the SFR in the core powering just the dark cone outflow were
0.2  \msun\ yr$^{-1}$ instead of
0.1  \msun\ yr$^{-1}$ used above, then
with the same arguments as above, $v_z$ would be 23 \kms\
 on the dark cone.  With $v _{\rm outflow}$ = 13 \kms,
this gives $i_{\rm feati}$ = $55\degr$, and, assuming the warp has the same tilt axis as the
central disk of NGC 2207,
 a galaxy warp at Feature i of $55\degr\ - 35\degr$ =
  $20\degr$.  Then  for the dark cone, $H_z$ at $d$ = $- 2''$ would be 400 pc, and
 the time required for the molecular gas to travel from $d$ = $-1''$ to
$d$ = $-2''$ with a constant speed $v_z$ of 23 \kms\ would decrease to 9 Myr.

\section{Possible Origin of Feature i}

Since Feature i is an unusually active region, the question is how did it originate. The models by \cite{struck05}
for the encounter between NGC 2207 and IC 2163 did not produce the spiral arms of NGC 2207. As
discussed in some detail in that paper, this is because the encounter is retrograde with respect to
NGC 2207, and thus the cumulative tidal forces on NGC 2207 are relatively weak. These models find
that pre-existing spiral arms in NGC 2207 are not destroyed by the interaction, but there are two
strong effects that can impact the arms.

In the primary model of  \citet{struck05}, IC 2163 approaches NGC 2207 on the western side and swings $180\degr$
around the outer edge of  the NGC 2207 disk to its present position  on the eastern side.
 As pointed out  by \citet{struck05},
a significant amount of mass is pulled off  IC 2163 and
captured onto the disk of NGC 2207.  However,
most of this mass transfer occurs in a short time interval about
halfway through the  encounter and falls  onto a different part of the disk than where
Feature i is subsequently located.  By the time  the future Feature i passes
closest to the transfer stream, the amount of mass
transferred is much smaller, so it seems unlikely to trigger the most active region in the disk.

\citet{struck05} did not mention a second effect, which arises from the fact that
IC 2163 pulls the nearest parts of the NGC 2207 disk backwards, against
the orbital flow. The loss of angular momentum compresses the disk. \citet{elmegreen95a}
find that, aside from the spiral arms, the radial brightness profile in {\it B} and {\it R}
on the western side of NGC 2207 is relatively flat for
 radial distances $40''$ to $90''$, corresponding to the 8 kpc-wide \HI\ ring,  and drops off
steeply beyond $90''$  (i. e., a little beyond Feature i).
This unusually flat radial profile with a sharp cutoff is evidence for the compression of the disk resulting from
loss of angular momentum.  The compression and associated velocity gradients can produce or enhance
clumpiness. In NGC 2207 five of the six very massive \HI\ clouds,  each with \HI\ mass in excess of
$10^8$ \msun,  lie on the western or eastern portions of the \HI\ ring/partial ring (see Figure\,\ref{fig1}).
Although these massive \HI\ clumps are likely to result, in part, from the bead instability of rings, the radial
gradient in the angular momentum loss may have played a role in increasing their mass, either directly or
by increasing the \HI\ velocity  dispersion and thus the gravitational Jeans mass.
Feature i lies on the spiral arm somewhat beyond the  massive  \HI\ clouds N2 and N3
 (see Figure\,\ref{fig1}).   We speculate that at the location where Feature i
subsequently developed, there was a large \HI\ clump on the pre-existing spiral arm,  and  the
radial gradient in angular momentum loss created an inward crashing gas stream from the outer
disk that hit the spiral arm at this clump as
the disk of NGC 2207  rotated clockwise.
The two intersecting flows
enhanced the mass and density of the clump and triggered the starburst that became Feature i.

\section{Conclusions}

New ALMA \CO, HCO$^+$, and 100 GHz continuum observations and Gemini NIFS K-band spectra are presented
and combined  with previous {\it  HST} optical, radio continuum, {\it Spitzer}, and \HI\ data to study
Feature i, a starburst clump on an outer arm of the interacting galaxy NGC 2207.
 Feature i has one of the  highest star formation rates
among the set of 1700 star forming complexes in interacting galaxies or
normal spirals measured by \citet{smith16}.
In its grazing collision with IC 2163, NGC 2207 suffered a retrograde encounter, which caused loss of
angular momentum and compression of its outer disk.  The double compression from the inward flowing gas
intersecting with an \HI\ clump on the pre-existing spiral arm may have produced Feature i.

One of the two CO emission lobes in  Feature i coincides with
an optically-opaque dust cone extending a plane-of-sky distance $\sim 500$ pc
from the center of its 170 pc core.   Our \HI\ data have too low a spatial resolution
for identifying this structure in \HI.
The CO column density can account for the dark cone extinction beyond the core
only if almost all the gas
and dust along the cone is on the near side (relative to us) of the NGC 2207 disk. In
 CO we find two outflows of cold molecular gas,
which we assume is streaming  perpendicular to the galaxy disk:
a) along the dark cone an approaching outflow
(consistent with the extinction argument) with velocity perpendicular to the
disk $v_z \sim 16$ \kms\ and an
estimated molecular mass $ 8 \times 10^6$ \msun, and b) on the opposite side, a receding outflow
(also detected in HCO$^+$) at $v_z  \sim 28$ \kms, but less extensive and of unknown mass.
For the dark cone, the  kinetic energy of bulk motion, $\sim 2 \times 10^{52}$ ergs,
can be supplied by supernovae and the stellar winds of high mass stars in the core  if
 star formation at the observed rate is integrated over 10 Myr.
 If the NGC  2207 disk is not warped at Feature i, then
the outflow along the dark cone is detected in CO to a height of $\sim 0.9$ kpc above the disk.

The core contains a $2 \times 10^6$ \msun\ star complex with
an age of 0.6 Myr. The measured Br$\gamma$ flux of the core is consistent with that expected from
this complex without requiring a top-heavy or otherwise unusual IMF.
The Br$\gamma$ and  ro-vibrational H$_2$  lines have  FWHM linewidths of  82 \kms\ and 64 \kms, respectively,
 probably a
combination  of  high turbulence plus some expansion of the ionized gas. The symmetry of the
Br$\gamma$ line profile  provides an upper limit of 22 \kms (the instrumental line-width) to
any (asymmetric) outflow of ionized gas.

In 2001, the \sixcm\ radio continuum flux density of the core was nearly double that in 1986, indicative of
either a radio SN or some other type of energetic outburst. The possible ULX detected by \citet{mineo14}
is nearly coincident with the radio continuum peak. Because of the very high extinction at that location,
the measured X-ray flux of the ULX could be consistent with an intermediate mass black hole.
Short-term radio variability associated with an intermediate mass black hole would make it easier to
reconcile the measured Br$\gamma$ flux with the \sixcm\ radio continuum flux density measured in 1986.

Our new observations combined with our own multispectral data set and data from the literature reveal Feature i to be
a heavily obscured region of intense activity on an outer spiral arm. Given the  $\sim 2 \times 10^8$ \msun\
 mass of cold gas in Feature i, its precusor may have been analogous to the massive ($10^8$ - $10^9$ \msun) \HI\ clouds found in regions of high \HI\ velocity dispersion  in
NGC 2207/IC 2163 and other galaxies undergoing close encounters  \citep{kaufman99}.
 Some of the usual sites of highly luminous  extra-nuclear star-forming clumps in large galaxies are
(a) the overlap regions in galaxy mergers, e.g., the Antennae \citep{gilbert07}, (b)
the hinge clumps produced by  the multiple converging flows near the base of a tidal tail in prograde
encounters, e.g., Arp 240   \citep{smith14}, and
 (c) the expanding ring of a collisional ring galaxy, e.g., the
Cartwheel galaxy \citep{higdon95}.
 Feature i illustrates that galaxies suffering close retrograde encounters  in which loss of angular momentum
by the outer disk causes disk contraction
 can also trigger the formation of  a highly luminous star-forming clump under suitable conditions, e.g., if
 the gradient in angular momentum loss by the outer  disk creates
an inward crashing stream that collides with a  massive \HI\ cloud as it is being
compressed  by a spiral density wave.

\acknowledgments

   This paper makes use of the following ALMA data:
   ADS/JAO.ALMA\#2012.1.00357.S and ADS/JAO.ALMA\#2013.1.00041.S.
 ALMA is a partnership of ESO (representing
   its member states), NSF (USA) and NINS (Japan), together with NRC
   (Canada), MOST and ASIAA (Taiwan) and KASI (Republic of Korea), in
   cooperation with the Republic of Chile. The Joint ALMA Observatory is
   operated by ESO, AUI/NRAO and NAOJ.
  The National Radio Astronomy Observatory is a facility of the National
  Science Foundation operated under cooperative agreement by Associated
   Universities, Inc.

This paper presents observations obtained at the Gemini Observatory under
program  GN-2016B-FT-26.  The Gemini Observatory is operated by
the Association of Universities for Research in Astronomy, Inc. under a cooperative
agreement with the NSF on behalf of the Gemini partnership: the
National Science Foundation (United States),  National Research Council (Canada),
CONICYT (Chile), Ministerio de Ciencia, Tecnologia e Innovaci\'{o}n Productiva
(Argentina), Minist\'{e}rio da Ci\^{e}ncia, Tecnologia e Inova\c{c}\~{a}o
(Brazil), and Korea Astronomy and Space Science Institute (Republic of Korea).

 This research made use of the NASA/IPAC Extragalactic
 Database (NED) which is operated by the Jet Propulsion Laboratory, California
Institute of Technology, under contract with the National Aeronautics and Space
Administration.

We are grateful to Kartik Sheth for his generous help on this project,
and to Stephanie Juneau for early discussions about this research.
 We thank Jonathan Westcott for his contribution in obtaining the re-calibrated
ALMA uv-data for Cycles 1 and 2 and the ALMA staff for doing the recalibration.
  We thank the referee for making detailed
suggestions to improve the presentation of our results.

E.B. acknowledges support from the UK Science and Technology
Facilities Council [grant number ST/M001008/1].


\end{document}